\documentclass[10pt,journal,compsoc]{IEEEtran}
\usepackage[utf8x]{inputenc}
\usepackage[T1]{fontenc}

\usepackage{cite}
\usepackage{booktabs}   
\usepackage{subcaption} 
\usepackage{tabularx}
\usepackage{listings}
\usepackage{xcolor}
\usepackage[bookmarks=false,colorlinks,linkcolor=blue,citecolor=blue,urlcolor=blue]{hyperref}
\usepackage{cleveref}
\usepackage{tabularx}
\usepackage{booktabs}
\usepackage{mdframed} 
\usepackage{verbatim} 
\usepackage{paralist}
\usepackage{listings}
\usepackage{courier}
\usepackage{xspace}
\usepackage{balance}
\usepackage{algorithm} 
\usepackage{algorithmic}
\usepackage{multirow}
\usepackage{framed}
\usepackage[]{graphicx}
\usepackage{caption}
\usepackage{subcaption}
\usepackage[table]{colortbl}

\newcolumntype{s}{>{\hsize=.75\hsize}X}
\newcolumntype{m}{>{\hsize=1.25\hsize}X}
\newcommand{\TODO}[1]{\textcolor{red}{#1}\GenericWarning{}{LaTeX Warning: TODO: #1}}\newcommand\todo\TODO

\usepackage{graphicx} 
\usepackage[font=small,skip=0pt]{caption}
\usepackage[export]{adjustbox}
\usepackage{tikz}
\usepackage{lipsum}

\lstdefinestyle{myCustomMatlabStyle}{
  columns=fullflexible,
  stepnumber=1,
  numbersep=10pt,
  tabsize=1,
  showspaces=false,
  showstringspaces=false
  escapeinside={(*@}{@*)},
  basicstyle=\footnotesize,
}

\usepackage{listings}
\lstdefinelanguage{diff}{
  language=java,
  basicstyle=\ttfamily\scriptsize,
  sensitive=true,
  morecomment=[f][\color{gray}][0]{diff},
  morecomment=[f][\color{gray}][0]{index},
  morecomment=[f][\color{blue}][0]{@@},
  morecomment=[f][\color{magenta}][0]{***},
  morecomment=[f][\color{violet}][0]{!},
  morecomment=[f][\color{red!60!black}][0]{-},
  morecomment=[f][\color{green!60!black}][0]{+},
  morecomment=[f][\color{magenta}][0]{---},
  morecomment=[f][\color{magenta}][0]{+++},
  morecomment=[f][\color{gray}][0]{Binary},
  morecomment=[f][\color{gray}][0]{Only},
  morecomment=[f][\color{gray}][0]{old},
  morecomment=[f][\color{gray}][0]{new},
  morecomment=[f][\color{gray}][0]{rename},
  morecomment=[f][\color{gray}][0]{similarity},
  morecomment=[f][\color{gray}][0]{deleted},
  morecomment=[f][\color{magenta}][0]{***************},
  morecomment=[f][\color{red!60!black}][0]<,
  morecomment=[f][\color{green!60!black}][0]>,
  morecomment=[f][\color{blue}][0]{0},
  morecomment=[f][\color{blue}][0]{1},
  morecomment=[f][\color{blue}][0]{2},
  morecomment=[f][\color{blue}][0]{3},
  morecomment=[f][\color{blue}][0]{4},
  morecomment=[f][\color{blue}][0]{5},
  morecomment=[f][\color{blue}][0]{6},
  morecomment=[f][\color{blue}][0]{7},
  morecomment=[f][\color{blue}][0]{8},
  morecomment=[f][\color{blue}][0]{9},
  }[comments]

\begin{document}

\title{Characterizing the Usage, Evolution and Impact of Java Annotations in Practice}

\author{Zhongxing~Yu,
        Chenggang~Bai,
        Lionel~Seinturier
        and Martin~Monperrus
\IEEEcompsocitemizethanks{\IEEEcompsocthanksitem Z. Yu and M. Monperrus are with KTH royal institute of technology.
E-mail: zhoyu@kth.se and martin.monperrus@csc.kth.se
\IEEEcompsocthanksitem C. Bai is with Beihang University.
E-mail: bcg@buaa.edu.cn
\IEEEcompsocthanksitem L. Seinturier is with Inria Lille Nord Europe and University of Lille. E-mail: Lionel.Seinturier@inria.fr}
\thanks{Manuscript received ...; revised ....}}

\markboth{TO APPEAR IN IEEE TRANSACTIONS ON SOFTWARE ENGINEERING}%
{Yu \MakeLowercase{\textit{et al.}}}

\IEEEtitleabstractindextext{%
\begin{abstract}
Annotations have been formally introduced into Java since Java 5. Since then, annotations have been widely used by the Java community for different purposes, such as compiler guidance and runtime processing. Despite the ever-growing use, there is still limited empirical knowledge about the actual usage of annotations in practice, the changes made to annotations during software evolution, and the potential impact of annotations on code quality. To fill this gap, we perform the first large-scale empirical study about Java annotations on 1,094 notable open-source projects hosted on GitHub. Our study systematically investigates annotation usage, annotation evolution, and annotation impact, and generates 10 novel and important findings. We also present the implications of our findings, which shed light for developers, researchers, tool builders, and language or library designers in order to improve all facets of Java annotation engineering. 
\end{abstract}

\begin{IEEEkeywords}
Annotation, Software Evolution, Empirical Study, Statistical Modelling.
\end{IEEEkeywords}}

\maketitle
\IEEEdisplaynontitleabstractindextext
\IEEEpeerreviewmaketitle
\ifCLASSOPTIONcompsoc
\IEEEraisesectionheading{\section{Introduction}\label{sec:introduction}}
\else
\section{Introduction}
\label{sec:introduction}
\fi

In programming languages, annotations are constructs for declaratively associating additional metadata information to program elements. In Java, they have been formally introduced since Java 5 \cite{JLS3}. 
The additional metadata information can be used for different purposes, such as the three possible uses pointed out by Java official tutorial \cite{tutorial}: guidance for the compiler, compile-time or deployment-time processing, and runtime processing. Besides, it is also widely known that many industry frameworks such as Spring, Hibernate, and JUnit use annotations to customize program behaviours. In academia, there also exist tools or frameworks which make use of compile time annotation processing to enforce stronger type checking \cite{isstatypechecker} or generate code \cite{OOPSLA15}. 

Despite the massive use of annotations among the Java community since its introduction, there is still limited empirical evidence about the actual usage of annotations in practice (\textit{annotation usage}), the changes made to annotations during the life-cycle (\textit{annotation evolution}), and the actual impact of using annotations on code quality (\textit{annotation impact}). Studying these questions is essential for researchers to better understand the  Java annotation language feature, for developers to improve their annotation skills, for language or library designers to improve their annotation designs, and for tool builders to come up with tools for helping developers use annotations. 

In this paper, we perform a large-scale empirical study about Java annotation. To the best of our knowledge, this study is the first to be really large scale, based on a corpus of 1,094 notable open-source projects hosted on GitHub. For studying \textit{annotation usage}, we investigate the density of annotations in the code, what kinds of developers use annotations, and what types of values are used to configure annotations. For studying \textit{annotation evolution}, we separate \textit{code dependent annotation changes} (changes that are related with other code change) from \textit{code independent annotation changes} (changes that are not related with other code changes), and study the characteristics of different kinds of \textit{code independent annotation changes}. For studying \textit{annotation impact}, we use regression model to explore the correlation between use of annotations and code quality. 

Our large scale study enables us to identify 10 major findings that have important implications for software engineering. These findings and their implications are summarized in \autoref{tab:finding}. 

\begin{table*}
 \caption{Our Major Findings And Their Implications Based on Studying Java Annotations in 1,094 Open-source Projects.}
  \label{tab:finding}
  \centering
  \footnotesize
  \begin{tabular}{|p{8.5cm}|p{8.5cm}|}
    \hline
     \cellcolor{gray!55} \textbf{Annotation Usage} (Section 4)& \cellcolor{gray!55} \textbf{Implication} \\
    \hline
   (1) In our corpus, the median value of annotations per project, per LOC (file unit), and per annotated program element is 1,707, 0.04, and 1 respectively, but the maximal values can be extremely large for all the three metrics. In particular, some developers tend to use heavily repeated annotations to encode processing logic for complex conditions. & (1) Using annotations is pervasive in practice and annotations are used reasonably most often, but can be overused occasionally (such as using repeating annotations heavily). IDE can warn developers if annotations are overused, and the numerical data summarized in this paper and detailed online can be referenced to set the thresholds. \\
   \hline
   (2) There exists a strong relationship between using annotations and developer ownership, and developers with high ownership of a file are more likely to use annotations. & (2) Using annotation requires good understanding of both the annotation and the annotated code. Better tutorials and dedicated training would help novice developers (with low ownership) to more use annotations. \\
    \hline
    (3) Nearly three-quarter (73.71\%) of the assigned annotation values are typed as String, and the String content could actually be better typed (for instance, as Class or Primitive) in some cases. One important reason is that some annotation members are wrongly designed as String type. & (3) Developers could better benefit from static type checking with better annotation design. This calls for better support in IDEs to identify bad smell in annotation design and to suggest appropriate refactorings. \\
    \hline
    \hline
    \cellcolor{gray!55} \textbf{Annotation Evolution} (Section 5)& \cellcolor{gray!55} \textbf{Implication}  \\
    \hline
    (4) Annotations are actively maintained. Most annotation changes are consistent changes together with other code changes, but there also exists a non-neglectable percentage (5.39\%-13.43\%) of annotation changes that are independent of other code changes. & (4) The current practice of using annotations is not good enough and some developers tend to use annotations in a subjective and arbitrary way, introducing annotation problems. Developers take efforts to address the problems as after-thoughts, calling for dedicated tool that can support the process of systematic annotation testing. \\
    \hline
    (5) For annotation add changes, three major reasons are clearing IDE warning (31.5\% ), clarifying specific facts about usage (33.6\%), and introducing new feature (25.5\%). There also exist some annotation move changes, which inconsistently move between outer and nested program elements. & (5) As after-thoughts, developers actively add annotations to the existing program elements to improve code readability and quality. When the annotations are effective, there sometimes exists a confusion about whether to put an annotation on the outer program element or the nested program element. \\
    \hline
    (6) For annotation deletion changes, three major reasons are inconsistent (31.5\%), redundant (15.4\%), and wrong (24.5\%) annotations. Some redundant annotations can (easily) be identified by checking the attribute of the annotated program element. Most wrong annotations are related with nullability analysis, dependency injection, mapping and data binding, and concurrency. & (6) Tools can do static analysis to detect annotation bad smells: for instance some redundant annotations can be found with specific templates about unnecessary usage scenarios. Annotation testing is important, in particular for those kinds of annotations that can easily get wrong according to our finding. \\
    \hline
   (7) For annotation replacement changes, 32.3\% of them are switching to opposite annotations and 19.79\% of them are switching to "same name" annotations from other libraries. & (7) There exists the prospect of automatically detecting and repairing annotation bugs that use opposite annotations of the correct ones, and unified standardization for popular annotations with different API providers will be beneficial. \\
   \hline
   (8) For annotation value update changes, four typical update behaviors are overriding or restoring default value (28.4\%), improving string of static text (23.9\%), correcting string of code (11.7\%), and correcting string of class name (10.4\%). & (8) Testing annotations should pay special attention to annotation members with default values (especially boolean type members) and String type annotation members that contain code as the content. Advanced IDE could also highlight potential syntax problems with code used as (String type) annotation value.\\
   \hline
   (9) For code independent annotation changes, most of them (81.78\%) change in the "Group Change" manner, i.e., multiple exactly same annotation changes happen in a single commit. & (9) We envision annotation change recommendation tools to help developers with the annotation change process. For instance, by analyzing some manually finished changes or recent changes, they could recommend the remaining changes to be done.\\
    \hline
    \hline
    \cellcolor{gray!55} \textbf{Annotation Impact}(Section 6) & \cellcolor{gray!55} \textbf{Implication}  \\
    \hline
    (10) There exists a small but significant relationship between annotation uses and code error-proneness: Java code with annotations tends to be less error-prone. The relationship is larger for bigger files as well as for files with fewer developers and commits. &  (10) Java annotations in general should be encouraged to use as it can potentially lead to an improvement in software quality. \\
    \hline
\end{tabular}
\end{table*}

Overall, the contributions of this paper are: 

\begin{itemize}

\item 
We performed the first large-scale and systematic empirical study about Java annotation on 1,094 notable open-source Java projects hosted on GitHub.

\item We presented 10 novel and important empirical findings about Java annotation usage, annotation evolution, and annotation impact. We also presented 10 implications from our findings to help developers, researchers, tool builders, and language or library designers involved in annotation engineering.

\item We made the collected dataset publicly available for future research on this topic (\url{https://github.com/zhongxingyu/Java-Annotation-Study}).

\end{itemize}

The remainder of this paper is structured as follows. We first present background about Java annotation in Section 2. Section 3 explains the methodology for data collection and analysis. In Section 4, Section 5, and Section 6, we give detailed results of our study about annotation usage, annotation evolution, and annotation impact respectively. Section 7 discusses threats to validity, followed by Section 8 which gives some closely related work. Finally, we conclude this paper in Section 9.

\section{Background}
\label{sec:background}
\subsection{Motivation for Annotation}
The need for metadata information about source code in Java has been around for quite a while, and the Java platform provides some ad-hoc annotation mechanisms to provide metadata traditionally. For instance, the marker interface and the @\texttt{deprecated} javadoc tag have both been widely used by developers. While these existing ad-hoc annotation mechanisms are useful, they sometimes do not generalize and scale well. To formalize and standardize the use of metadata in Java, Java 5 formally introduced annotation as a syntactic feature \cite{JLS3}.

\subsection{Annotation Definition and Use} 
Before the use of an annotation, it should first be declared. The declaration of an annotation type is similar to the declaration of an interface type, and the compilers use keyword @interface to recognize it. \autoref{lst:annotationformat} shows the format of a typical annotation type declaration, which mainly includes use of some meta-annotations and declaration of an annotation type body.

\begin{figure}[t]
\begin{lstlisting}
 //meta-annotations, e.g., @Retention, @Target
 <modifiers> @interface <annotation-type-name> { 
 //annotation type body 
 method(i):<return type><method name><default value>
 }
\end{lstlisting}
\caption{Format of Annotation Type Declaration in Java.}
\label{lst:annotationformat}
\end{figure}

Meta-annotations are annotation types designed for annotating other annotation type declarations, and built-in meta-annotations which define the basic property and usage scenario of the declared annotation type are typically used. For example, meta-annotation @\texttt{Target} is used to specify the targeted kinds of program elements for an annotation type. Initially, annotations are only allowed to annotate the program element declarations, but the application on type use is allowed since Java 8 \cite{Java8}. For another example, meta-annotation @\texttt{Retention} determines at what point annotations should be discarded. The annotations can be kept only on the source file, or kept on the binary class file, or even available at runtime. 

The annotation type body can be empty or contain a set of method declarations. Each declared method, if any, is a \emph{member} of the declared annotation type and should not have any parameters or a throws clause. In addition, the method can possibly have default values and the return types should only be primitives, String, Class, enums, annotations, and arrays of the preceding types \cite{JLS3}.

Developers can use annotations declared by the Java platform (called built-in annotations), third-party libraries or declare annotations themselves (called custom annotations). To use an annotation in the code, it should be put on program elements allowed by this annotation and the value given to the annotation, if exists, takes the form of method-value pairs. Besides, the value given to each method must be a compile time constant. For example, \autoref{lst:built-in annotationoverride} shows the declaration and use case for the built-in annotation @\texttt{Override}. According to the meta-annotation @\texttt{Target}, @\texttt{Override} can only be used to annotate method declaration. For another example,  \autoref{lst:built-in annotationjunit} shows the declaration and use case for the JUnit framework annotation @\texttt{RunWith}. This annotation can be used to annotate type declaration according to the meta-annotation @\texttt{Target}, and has a method named \emph{value} which is embodied with constant \emph{SpringClassRunner.class} in this example. 

\begin{figure}[t]
\centering
\begin{lstlisting} 
@Target(METHOD)               @Override
@Retention(SOURCE)            public int hashCode(){
public @interface Override{#<-># //override code
}                             }
\end{lstlisting}
\caption{The Source Code and Use Case for the Built-in Annotation @Override.}
\label{lst:built-in annotationoverride}
\end{figure}

\begin{figure}[t]
\centering
\begin{lstlisting} 
@Target(TYPE)                 @RunWith(value=Spring
@Retention(RUNTIME)           ClassRunner.class) 
public @interface RunWith{    class SpringTests{
Class<?> value();          #<-># //test code
}                             }
\end{lstlisting}
\caption{The Source Code and Use Case for the JUnit Annotation @RunWith.}
\label{lst:built-in annotationjunit}
\end{figure}

\subsection{Annotation Processing} 
There are several different ways to process annotations so that they take effect. 

During the compile time, the Java platform provides a compiler-plugin mechanism to deal with annotations. An annotation processor for a specific annotation can be first defined and plugged to the compilation process, and then the compiler will inspect the codebase against the presence of the annotation and respond according to the processing procedure specified in the annotation processor. For example, the annotation processor for @Override will check whether the method \texttt{hashCode()} in \autoref{lst:built-in annotationoverride} is really an override method, and cause a compile error if not. If the annotations are not discarded after the compilation process and are kept in the compiled class files, the annotation metadata can be used for byte-code level post processing. Finally, for those annotations that are set to be available at runtime, reflection can be used to retrieve the annotation metadata and the program behaviour can be customized accordingly. For instance, according to the retrieved metadata for the @\texttt{RunWith} annotation in \autoref{lst:built-in annotationjunit}, \emph{SpringClassRunner} will be used to run the test cases instead of the default JUnit test runner.


\section{Experimental Methodology}
\label{sec:methodology}
 In this section, we describe our methodology to perform our original large-scale study of annotations in Java.
 
\subsection{Study Subjects}
We use large and popular Java projects hosted in Github as the study subjects. Our project selection process is as follows. We first retrieve all non-fork Java repositories whose stargazers counts (which indicate the popularity of projects) are larger than 100. Then, to avoid inactive projects, we discard projects that do not have a single event of activity within 6 months prior to our data collection date, October 28, 2017. This leaves us with 5,683 repositories. Next, following the guidance in \cite{Kalliamvakou:2016:ISP:2992358.2992445}, we exclude repositories with less than 3 contributors to further focus on the active projects. This leaves us with 4,035 repositories. For the remaining repositories, we prioritize them according to the number of commits. Finally, we use all the 1,094 repositories whose commit numbers are larger than 1,000 as the study subjects. The studied repositories include most well-known Java projects in Github such as android/platform\_frameworks\_base, apache/hadoop, spring\-projects/spring\-frameworks, and gradle/gradle, etc. These studied projects cover typical domains such as framework, library, database, middleware, and web utility, etc. 

\subsection{Data Collection}
\label{sec:datacollection}
To study annotation usage, annotation evolution, and annotation impact, we need to collect data about annotation uses in different file versions and the bug fixing history associated with each file. We next describe how we collect the data. As this paper focuses mainly on studying annotation use rather than annotation definition, we simply call annotation use as annotation in the latter sections of this paper unless otherwise specified.

\vspace{1.0mm}
\textbf{Retrieving and Comparing Annotations.} 
We use the Spoon library \cite{spoonlib} to retrieve annotations in a Java file. Spoon is a library that parses source code file into the AST and provides a set of APIs for conducting program analysis. In particular, it has a set of APIs for accessing annotations in the file. For each file, we thus use the APIs to get the annotations in the code.

To study annotation evolution, we need to compare the annotations in different versions of a file. For different file versions, the annotations can be added, deleted, and replaced. Besides, the changes can be just made to values of annotation members. Meanwhile, the annotations can be changed independently or in parallel with the annotated program elements. Consequently, we define and calculate the following 6 kinds of changes. \autoref{lst:change example} gives an illustration for each of these change kinds. 

\begin{itemize}
\item \textbf{ADD\_ALL:} The new file version adds a new annotation and the annotated program element does not exist in the old file version (i.e., also newly added).
\vspace{0.6mm}

\item \textbf{ADD\_ANN:} The new file version adds a new annotation and the annotated program element already exists in the old file version.
\vspace{0.6mm}

\item \textbf{DEL\_ALL:} The new file version deletes an existing annotation and the annotated program element does not exist in the new file version (i.e., also deleted).
\vspace{0.6mm}

\item \textbf{DEL\_ANN:} The new file version deletes an existing annotation and the annotated program element still exists in the new file version.
\vspace{0.6mm}

\item \textbf{CHANGE:} The annotation on a program element has been changed to a different annotation for the same program element in the new file version.
\vspace{0.6mm}

\item \textbf{UPDATE:} The annotation on a program element is also on the same program element in the new file version, but the annotation values are different.
\end{itemize}

\begin{figure}[t] 
\centering
\begin{lstlisting}
//version 1 (old version)  //version 2 (new version)
public class Example {       public class Example {
 !@GuardedBy("lock") !~(DEL_ANN)~ 
 Object field; #<------------># Object field;
                   ~(ADD_ANN)~!  @Override!
 void Method1() { #<---------># void Method1() {
 }                    ~(UPDATE)~}
!@SuppressWarnings("all")!#<-->#!@SuppressWarnings("nls")!
 void Method2() {   ~(CHANGE)~  void Method2() {
  !@Nullable! #<----------------># !@NotNull!
  String localVar1;            String localVar1;
  ...                          String localVar2;
 }                             String localVar3;
 !@Deprecated!~(DEL_ALL)~          !@NotNull!~(ADD_ALL)~
 void Method3(int param) {     String localVar4; 
 }                            }
}                            }
\end{lstlisting}
\caption{Examples of Different Kinds of Annotation Changes.}
\label{lst:change example}
\end{figure}

We use GumTree \cite{gumtree} to determine whether an annotated program element exists in different versions of a file. GumTree is an off-the-shelf, state-of-the-art tree differencing tool that computes AST-level program modifications. The manual evaluation has shown that GumTree produces good outputs that can well explain code changes for 137/144 (95.1\%) file pairs, and a large-scale automatic evaluation also shows that it is more accurate than other tree differencing tools in disclosing the essence of code changes \cite{gumtree}. We customize GumTree to output the changes to annotated program elements for different file versions. Note that the annotated program elements can also be updated or moved across different file versions, but we find these cases are relatively rare (see Section 5.1 for related statistics). To simplify, the annotated program element is considered to exist in both versions if it is updated or moved between two file versions in the above definition. In addition, an annotation can move from one program element to another program element. However, we do not explicitly define annotation move changes for two reasons. First, they are rare in practice (see Section 5.2.1 for related statistics). Second, it is difficult to establish the circumstance under which we can view the Gumtree detected annotation move changes as real semantic annotation moves. To achieve this, we would need to manually consider the semantics of both the annotation itself and the annotated code. For instance, suppose GumTree detects that a specific annotation \emph{ann} on a parameter of method \emph{a} has moved to a local variable of method \emph{b}, yet this does not necessarily mean that this is an annotation move change. 

\textbf{Retrieving Project Evolution History.} 
For each source file, our study process needs to parse its different versions and do comparisons. As there will be too many file parsings and parsing result comparisons if we consider the complete evolution history for each project, we thus focus on the evolution history of the past three years to make the time manageable. To download the evolution history of the past three years, we use the git log command \texttt{git log --after="2014-10-28" --before="2017-10-28" --no-merges --numstat}. The command will output all non-merged commits along with the commit log, author date, author name and files that have been changed for each commit. For 21 projects, developers frequently commit to them and the size of the log file from the above command is larger than 15MB, we thus limit the considered commit history to the past one year for them.

Then, we remove changes made to non-java files by checking whether a changed file has the suffix ".java". Meanwhile, we also remove changes made to test files as we focus on annotations in source files in this study. The use of annotations in test files can be different from their uses in source files, and studying its characteristics is out of scope of this paper. To achieve this, we view a file as a test file in case the keyword "test" is present within the file name. For each remaining file changed in a certain commit, we compare annotations in the file version corresponding to this commit with that in the file version corresponding to the previous commit. 

\textbf{Collecting Bug-fixing Commits.} 
To get the bug-fixing commits, we use the method proposed by Mockus et al. \cite{icsmbug} to look for error related key words in commit message associated with each commit. First, we use Stanford natural language processing (NLP) tool \cite{nlptool} to convert each commit message to a bag-of-words and lemmatize the bag-of-words. Then, a commit is marked as a bug-fixing commit if the corresponding lemmatized bag-of-words contains at least one of the error related keywords: "error", "bug", "fix", "issue", "fixup", "bugfix", "npe", "mistake", "blunder", "incorrect", "fault", "defect", "flaw", "glitch", "gremlin", and "erroneous". Several previous studies \cite{commitclassify1}\cite{commitclassify2}\cite{commitclassify3}\cite{commitclassify4} have adopted the same method to identify bug-fixing commits. 

As this method in principle is a heuristic, we randomly select 500 commits \footnote{For the top 100 projects in terms of commit number, we randomly select 5 commits for each project.} and manually examine the commit messages to evaluate the accuracy of the above classification. If we are not sure from the commit message, we further examine the source code and linked issues, if exist, to help us understand the purpose of the commit. For 5 commits, we are still not able to determine whether they are bug-fixing commits or not. Out of the remaining 495 commits, 470 have been classified correctly (95\%) and 25 commits (5\%) have been classified incorrectly—16 false negatives and 9 false positives. Overall, the accuracy of the approach is considered acceptable.

\subsection{Statistical Methods}
\label{sec:regressionmodel}
Our study involves a large amount of data, thus we seek to use statistical methods to analyze the data.

First, boxplots are used to visually display the distributions of different populations. The visual results are further confirmed using the non parametric Wilcoxon test, and the test results are interpreted using the \textit{p}-values, which indicates the probability of a hypothesis being true by chance. 

Second, to evaluate annotation impact, we use regression model to establish the correlation between annotation and the occurrence of faults in the history of a file. The response variable in the model is \textbf{nbbug}, which is the number of estimated bug-fixing commits associated with a file during the commit history we considered. Our interested experimental variable is \textbf{nbAnnotation}, which we define as the average number of annotations across all the versions of a file during the considered commit history. For instance, suppose a file has 3 versions during the 3 years of commit history, and the first, second, and third version has 10, 14, and 18 annotations respectively, then the value of \textbf{nbAnnotation} for this file will be 14. When using regression models, it is important to control the confounding factors. Following previous studies \cite{commitclassify1, commitclassify3, premregressionmodel}, we use the following 3 control measures, 

\vspace{1mm} 
\noindent
\textbf{codesize}: the average size of the file (in terms of LOC) across all the versions of the file during the considered commit history. Previous study \cite{classsize} has shown that the size of the file is strongly correlated with the occurrence of defects in the file. In our model, we log-transform this control measure as it exhibits a log-linear distribution and is not strictly a count variable, and it has been shown that doing so can stabilize the variance and improve model fit \cite{commitclassify3,modelimpact}.

\vspace{1mm} 
\noindent
\textbf{nbCommit}: the number of commits made to the file during the considered commit history. In general, it has been shown that the more commits made to a file, the more likely errors can be introduced \cite{icse05codechurn}. . 

\vspace{1mm} 
\noindent
\textbf{nbDev}: the number of developers that have made commits to the file during the commit history we considered. Earlier study has shown that the number of developers has an impact on the delivered quality of the software \cite{ownership1}. 

\vspace{1mm} 

When establishing the regression model, a special challenge arising for our data is that a large number of files do not have any bug-fixing commits during the commit history we considered, i.e., the number of zero values overwhelms the non-zero values. If we fit a single regression model to the entire data collected, the results may be biased as it implicitly assumes that both the zero-defect and non-zero defect data come from the same distribution, which can be untrue. 

To deal with this issue, we use hurdle regression model \cite{hurdlemodel}, which helps handle data with excess zeros and over-dispersion. The hurdle regression model is a two-part model that specifies one process for zero counts (called hurdle model) and another process for positive counts (called count model). For our case, the hurdle model models the effect of passing from zero defect count to non-zero defect count, and the count model models the effect of going from one non-zero count to another non-zero count. In general, the two models use nonlinear multiple regression with different linking functions. In this paper, we use the default setting, which uses binomial regression for the hurdle model and poisson regression for the count model.  

\section{Results on Annotation Usage} 
\label{sec:annotationuse}
In this section, we study several questions about annotation usage, including density of annotations, users of annotations, and assigned annotation values for annotation members.                      
\subsection{Density of Annotations}
\begin{figure}
\centering
\begin{subfigure}{.25\textwidth}
  \centering
  \includegraphics[width=1.0\linewidth]{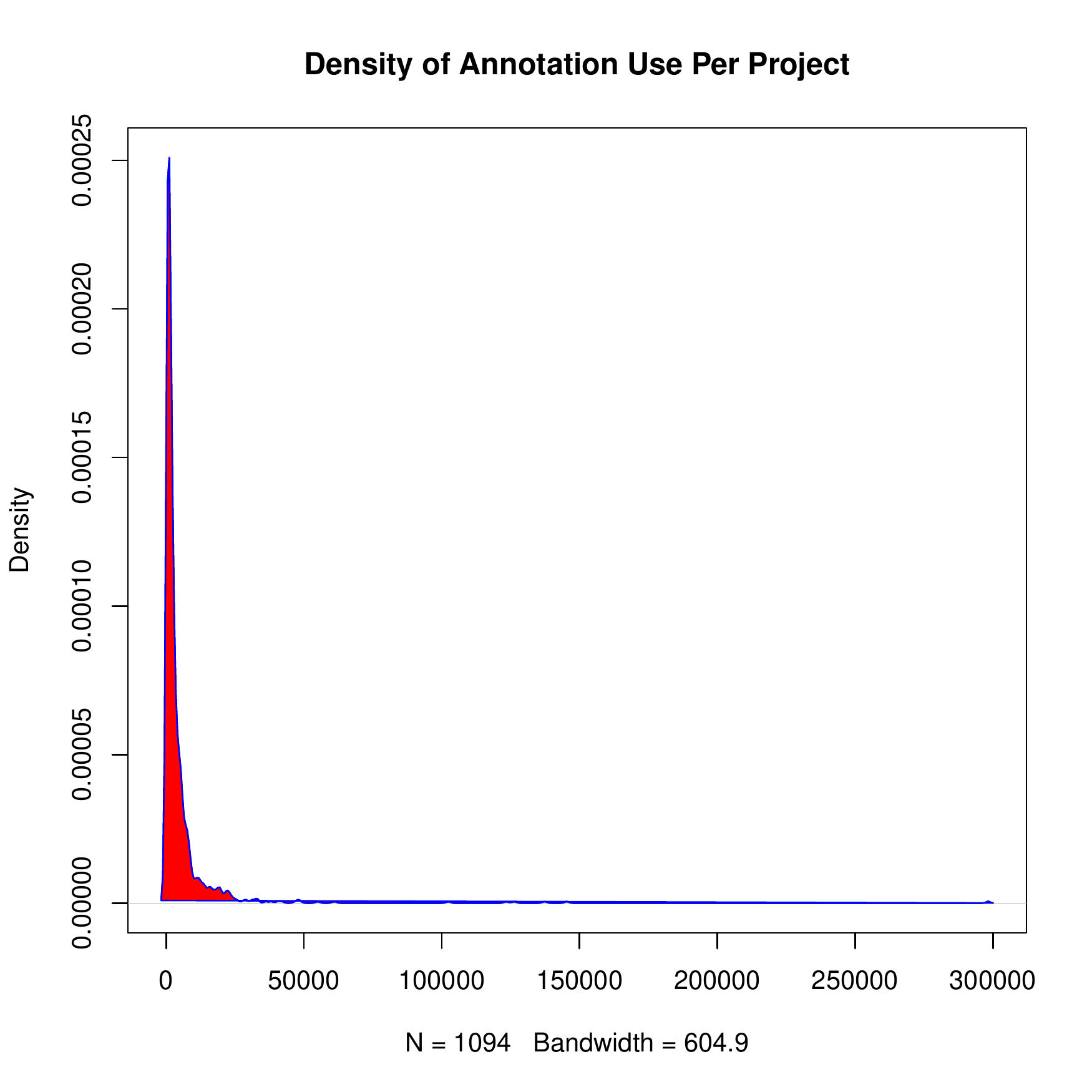}
  \caption{Annotation Use by Project}
  \label{fig:UseByProject}
\end{subfigure}%
\begin{subfigure}{.25\textwidth}
  \centering
  \includegraphics[width=1.0\linewidth]{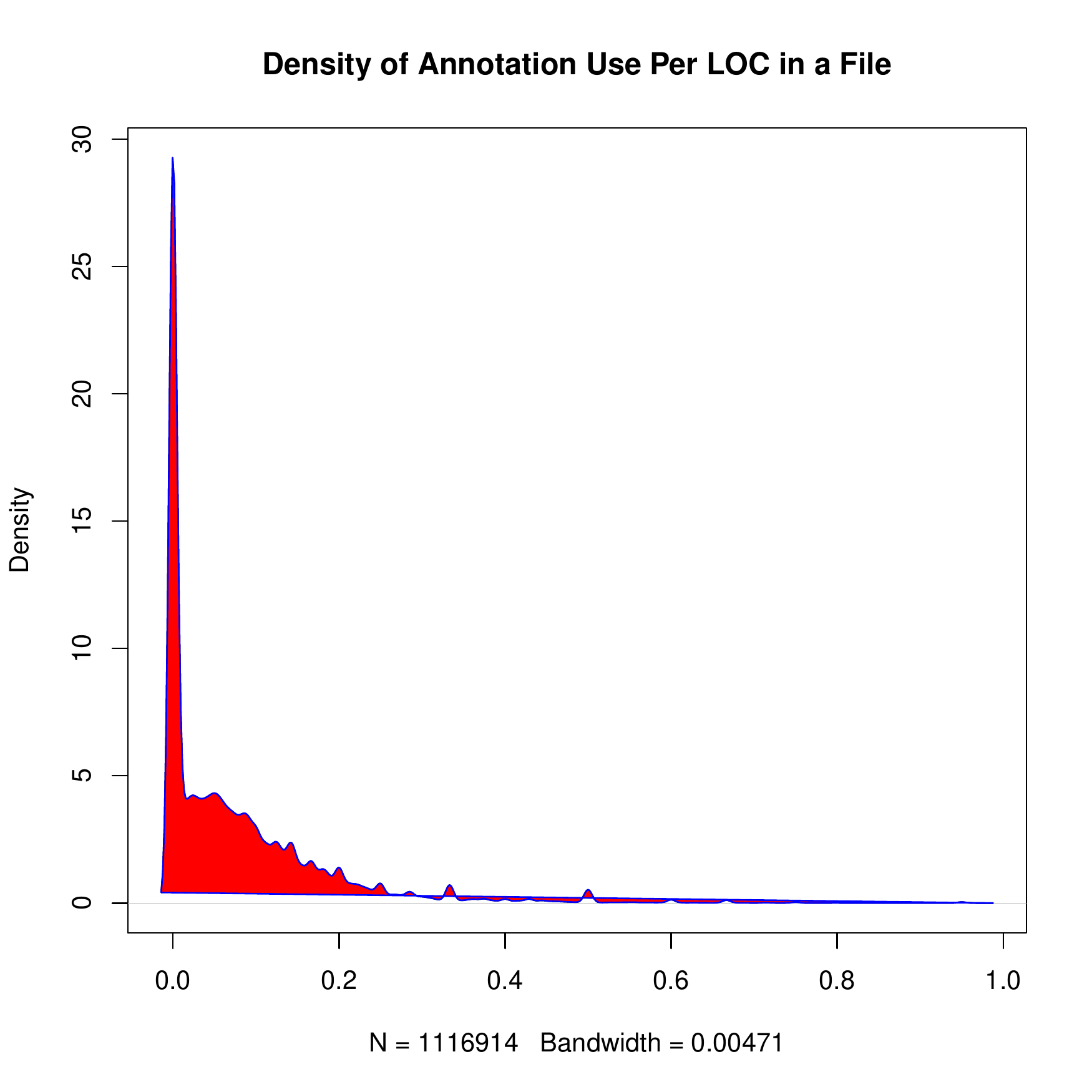}
  \caption{Annotation Use by LOC}
  \label{fig:UseByFile}
\end{subfigure}
\caption{Density of Annotation Use by Project and LOC in a File.}
\label{fig:UseProjectFile}
\end{figure}

We first want to see whether annotations are actually widely used by developers in practice. For the newest versions of the studied 1,094 projects, we observe that each of them has at least one annotation and there are 5,236,822 annotations in total. \autoref{fig:UseByProject} shows the density plot of the number of annotations for the 1,094 projects. It can be seen from the figure that most projects use a moderate number of annotations and a few projects use an extremely large number of annotations. More specifically, the minimum, 1st quartile, median, 3rd quartile, and maximum number of used annotations is 1, 763, 1,707, 4,413, and 298,284 respectively. Overall, we can conclude that the use of annotations is pervasive in modern software development process, which is in line with results of previous studies \cite{annotationicse, annotationseke}. 

A potential concern arises from \autoref{fig:UseByProject} is whether annotations will be overused in practice, which is widely considered bad as it can reduce the readability and even maintainability of the code \cite{annotationoveruse, annotationoveruseweb1, annotationoveruseweb2, annotationoveruseweb3}. To check this, we further calculate the value of annotation per line of code for each file, i.e., the number of annotations divided by the number of lines of code in a file. \autoref{fig:UseByFile} shows the density plot of the calculated value across all the 1,116,914 files for the studied 1,094 projects. We can see from the figure that the value is between 0.0 and 0.2 for most files, but can be larger than 0.5 or even close to 1.0 for some files. The minimum, 1st quartile, median, 3rd quartile, and maximum value is 0.0, 0.0, 0.04, 0.11, and 0.97 respectively. In particular, among the 1,116,914 files, the value is 0.0 for 380,770 files (34.1\%) and larger than 0.5 for 15,521 files (1.4\%). If the value is larger than 0.5, it implies that there are more annotations than other code for a certain file. From these data, we can see that while one third of the files do not use a single annotation and most of the files have a reasonable value of annotation per LOC, a few files have an extremely large value of annotation per LOC and thus suffer from the "annotation-overuse" problem.

In addition, we also calculate the number of annotations per annotated program element and \autoref{tab:annotationuseelement} shows the result. We can see that most annotated program elements have 1 or 2 annotations (4,409,525 cases, 98.8\%), but a few of them (2,978 cases) have more than 5 annotations. The maximal number of annotations on a single program element is 41, which is surely too extreme. These data again verify that annotations can be overused occasionally in practice.

In particular, we are surprised that a single program element can be annotated with so many annotations sometimes. To see the possible reason, we manually studied the 42 scenarios in \autoref{tab:annotationuseelement} for which a program element is annotated with greater than or equal to 10 annotations. We find that for 32 scenarios, a single annotation is used several times with different annotation values (note repeating annotations are allowed as of Java 8). Specifically, for 22 scenarios, more than 80\% of all annotation uses are in fact using a certain annotation with different annotation values. For these scenarios where a certain annotation is heavily repeated, annotations are typically used to encode processing logic for complex conditions, and each annotation value corresponds to the  processing logic for a specific condition. This kind of annotation usage is considered bad as real program logic is getting blurred by the heavily repeated annotations with different annotation values \cite{annotationoveruseweb1}. The numerical data summarized in this paper and detailed online can be referenced to set the thresholds for preventing heavily repeated annotations in particular, and annotation overuse in general. 

\begin{table}
 \caption{Number of Annotations per Program Element}
  \label{tab:annotationuseelement}
  \centering
  \footnotesize
  \begin{tabular}{|c|c||c|c|}
    \hline
     \# Annotations & Prevalence & \# Annotations & Prevalence \\
    \hline
    1 & 4,115,261 & 6 & 2,032\\
    2 & 294,264 & 7 & 687 \\
    3 & 35,382 & 8 & 156 \\
    4 & 10,280  & 9 & 61\\
    5 & 4,254 & $\geq$10 & 42 \\
    \hline
\end{tabular}
\end{table}

\vspace{0.6mm} 
\noindent
\textbf{Finding 1}: \textit{In our corpus, the median value of annotations per project, per LOC (file unit), and per annotated program element is 1,707, 0.04, and 1 respectively, but the maximal values can be extremely large for all the three metrics. In particular, some developers tend to use heavily repeated annotations to encode processing logic for complex conditions.} 

\vspace{0.6mm}
\noindent
\textbf{Implication 1}: \textit{Using annotations is pervasive in practice and annotations are used reasonably most often, but can be overused occasionally (such as using repeating annotations heavily). IDE can warn developers if annotations are overused, and the numerical data summarized in this paper and detailed online can be referenced to set the thresholds.}

\subsection{Users of Annotations}
\label{sec:userannotation}
Annotation is a relatively new language feature and it is not straight-forward to use it. It is reasonable to infer that developers who use annotations in general are familiar with the code and have a good conceptual understanding of the logic behind the code. Translating it into the concepts of collaborative software engineering \cite{ownership1}, the expectation is that for a specific file \textit{f}, developers who use annotations on it have a higher ownership of it than others who do not use annotations. We thus explore the relationship between ownership and using annotation, and see whether the relationship is as expected.

To explore the relationship, we first calculate the ownership of developers for each developer-file pair. Following the previous measure of ownership \cite{ownership2}, we calculate it as the percentage of changes made to a file by a certain developer. For instance, if there are 20 commits made to a specific file \textit{f} during its life-cycle and a single developer \textit{d} made 12 commits, then the ownership of \textit{d} for \textit{f} is 0.6. After calculating the ownership, for a certain file \textit{f}, we then separate all developers for it into two groups: one group contains developers who have used annotations and the other group contains developers who have not used annotations. If a developer has once added, deleted, or modified annotations (i.e.,the developer has made some of the 6 annotation change types defined in Section 3.2), this developer is considered to have used annotations. If any of the two groups is empty, we omit the file from our analysis. Finally, we calculate the median ownership of each group of developers for each file.

\begin{figure}
\centering
\begin{subfigure}{.25\textwidth}
  \centering
  \footnotesize
  \includegraphics[width=1.0\linewidth]{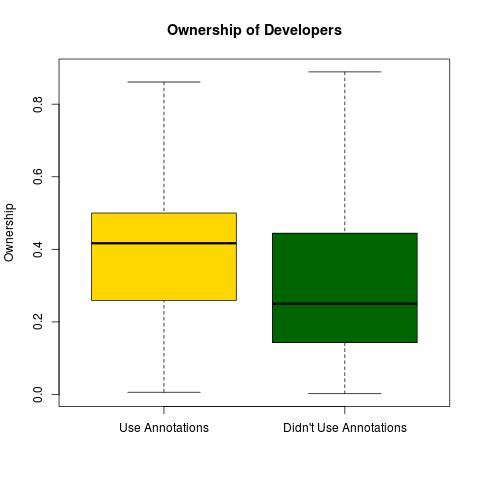}
  \caption{Ownership}
  \label{fig:ownership}
\end{subfigure}%
\begin{subfigure}{.25\textwidth}
  \centering
  \includegraphics[width=1.0\linewidth]{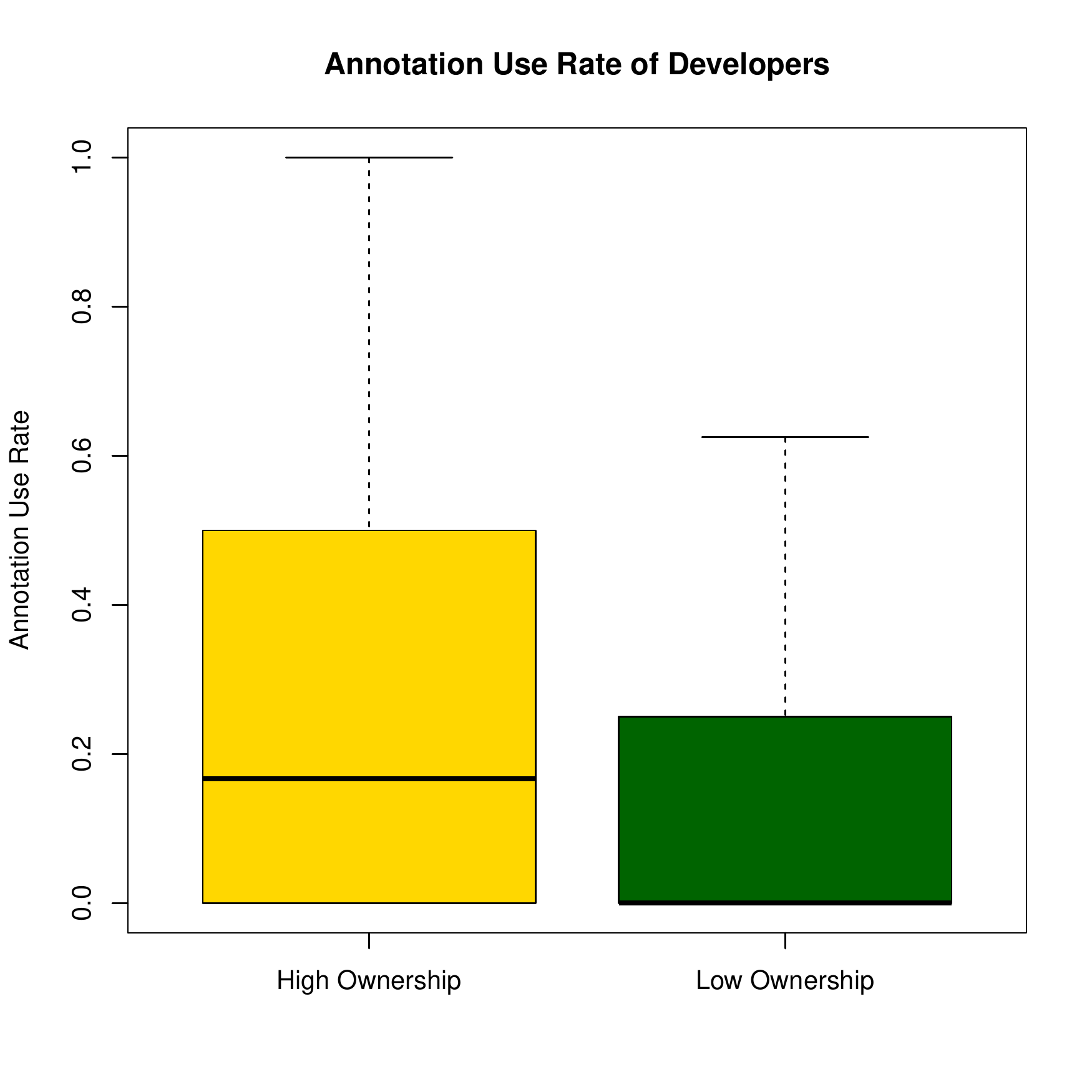}
  \caption{Annotation Use Rate}
  \label{fig:userate}
\end{subfigure}
\caption{Distribution of Ownership and Annotation Use Rate for Different Kinds of Developers.}
\label{fig:ownershipuserate}
\end{figure}

The result is shown in \autoref{fig:ownership}. From the boxplot, we can clearly see that developers who have used annotations are associated with higher ownership. The 1st quartile, median, 3rd quartile value of ownership for developers who have used annotations is 0.26, 0.42, and 0.50 respectively, whereas the corresponding value for developers who have not used annotations is 0.14, 0.25, and 0.44 respectively. The clear visual impression is confirmed by Wilcoxon test, which shows that the difference between the two groups is significant (\textit{p}-value less than 2.2*$e^{-16}$). Overall, the results support the expectation that developers who use annotations on the code have a higher ownership of it than others who do not use annotations. 

While developers who use annotations on a file have a higher ownership of it, this may simply because they make more commits to the file and not because the knowledge required for using annotations on the file. To further investigate this, for each developer \textit{d} of a certain file \textit{f}, we additionally calculate \textit{d}'s \textit{AnnotationUseRate} with regard to \textit{f}, defined as the number of commits that have used annotations across the total number of commits that \textit{d} has made to \textit{f}. For instance, if \textit{d} has made 10 commits to \textit{f} and 3 commits have used annotations, then the \textit{AnnotationUseRate} with regard to \textit{f} is 0.3 for \textit{d}. Then, we calculate the median ownerships for all developers of \textit{f} and separate the developers into two groups, one group contains developers whose ownership is larger than the median value and the other group opposite. If a file has a single developer or has not used any annotations during its life-cycle, we omit the file from our analysis. Finally, we calculate the average \textit{AnnotationUseRate} of each group of developers for each file. 

The result is shown in \autoref{fig:userate}. We can obviously see from the boxplot that developers with higher ownership have larger annotation use rate. The 1st quartile, median, 3rd quartile value of annotation use rate for developers with high ownership is 0.00, 0.17, and 0.50 respectively, and the corresponding value for developers with low ownership is 0.00, 0.00, and 0.25 respectively. The visual impression is again confirmed by Wilcoxon test. These results suggest that compared to developers with lower ownership of a file, developers with higher ownership of the file are more knowledgeable about using annotations in the file and consequently are much more likely to use annotations. 

\vspace{1.0mm} 
\noindent
\textbf{Finding 2}: \textit{There exists a strong relationship between using annotations and developer ownership, and developers with high ownership of a file are more likely to use annotations.}

\vspace{0.6mm}
\noindent
\textbf{Implication 2}: \textit{Using annotation requires good understanding of both the annotation and the annotated code. 
Better tutorials and dedicated training would help novice developers (with low ownership) to more use annotations.} 

\subsection{Annotation Values}
\label{sec:annotationvalue}
While some annotations do not have any annotation members (called marker annotations), others allow developers to explicitly specify values for annotation members so that more metadata information can be encoded. One important aspect pertaining to annotation uses is correctly specifying annotation values for the corresponding annotation members, otherwise unexpected program behavior may happen.

With regard to all the 5,236,822 annotations for the newest versions of the studied projects, annotation values have been explicitly specified for 965,777 annotations. \autoref{tab:annotationvalue} shows the summary of different types of assigned annotation values. In this table, the \emph{Value Type} column shows the type of assigned values, the \emph{Frequency} and \emph{Percentage} columns give the actual number and percentage for each value type respectively, and finally the \emph{\#Annotations} column shows the number of annotations that each value type is involved. Note that multiple values of the same or different types can be assigned to a single annotation. Besides, if the assigned value is an array of a certain type \emph{T}, it is deemed that a single value of type \emph{T} is used. We identity value types as follows: String values start with ", primitive values are matched with regular expression, Class values end with .class, 
and annotation values starts with @. It may happen that annotation values for these 4 value types are set with constant variables. If this happens, we are not able to know their exact types and distinguish them from enum types as more advanced whole program analysis is required. For these cases, we show them as "Unknown" in \autoref{tab:annotationvalue}. Our manual examination of a sample of 200 values with unknown types suggests that most (88.5\%) of the unknown types are String. 

\begin{table}
 \caption{Summary of Assigned Annotation Values.}
\label{tab:annotationvalue}
  \centering
  \small
  \begin{tabular}{|c|c|c|c|}
    \hline
  Value Type & Frequency & Percentage & \#Annotations \\
    \hline
    String & 861,463 & 62.57\% & 729,759 \\
    Primitive & 206,255  & 14.98\%  & 135,506\\
    Class & 83,174 & 6.04\% & 81,079 \\
    Annotation & 17,582 & 1.29\% & 16,310 \\
    \hline
    \hline
    Unknown & 208,190 & 15.12\%  & 186,695 \\
    \hline
\end{tabular} 
\end{table}

Among the identified clear value types, it can be seen from \autoref{tab:annotationvalue} that nearly three-quarter (73.71\%) of the assigned values are of type String. We randomly browse some String contents (i.e., the actual characters inside the double quotation marks "") and find that a non-small percentage of the String contents seem to be values of other types. As compilers cannot do type check for String contents, it is considered bad to assign an value as String if the value could actually be typed as primitive, Class, enum, or annotation. For instance, if developers assign a class name as String content (which in fact could be typed as Class), it is easy to make typo error when writing the name and moreover, the name could be changed due to code refactoring and it is common for developers to forget to update the String content of the name. Our result in Section 5.2.4 suggests that developers frequently make these mistakes. 

To investigate this problem, we use regular expressions for primitive value, annotation, and fully qualified Java class name (we confine it to have at least two dots) to match the String content and the matching result is given in \autoref{tab:stringmatching}. The \emph{Frequency} column gives the number that the String content matches with each matching type, and the \emph{\#Annotations} and \emph{\#Members} columns show the number of annotations and annotation members that each successful matching type is involved with respectively.
Even though the successful matching does not necessarily mean the String content is of a certain matching type and the annotation is wrong (as using annotation can contain rich context information), the large number of matched instances highlights the potential severity of the problem and this leads us to ask an question: are the types of some annotation members wrongly designed as String?

\begin{table}
 \caption{Summary of String Content Matching Result.}
\label{tab:stringmatching}
  \centering
  \small
  \begin{tabular}{|c|c|c|c|}
    \hline
  Matching Type & Frequency & \#Annotations & \#Members \\
    \hline
    Class Name & 18,716 & 490 & 582 \\
    Primitive & 14,142  & 309  & 348\\
    Annotation & 1,157 & 23 & 23 \\
    \hline
\end{tabular}
\end{table}

To answer the question, for the involved annotations in \autoref{tab:stringmatching}, we find the source files that define the annotations and read the comments for the involved annotation members to understand the usage scenario of the member and determine whether the designed member type String is correct. For matching type class name (582 annotation members) and primitive (348 annotation members) in \autoref{tab:stringmatching}, we randomly select 100 involved members respectively. \autoref{tab:manualchecking} gives a summary of the result. We can see from the column \emph{\#Wrong} that there are truly many wrongly designed String types. In particular, 49 ideally "Class" type annotation members and 23 ideally "Primitive" type annotation members have been wrongly designed as String type annotation members. \autoref{lst:wrongdefinitionexample} exemplifies these two kinds of wrongly designed annotation members.

We remark there is not too much bias for our manual check. The reason for the wrong design is obvious as the examples in \autoref{lst:wrongdefinitionexample} most of time, and we deem it as "Unclear" whenever we are not sure of the judgment. 
Besides, in theory, it can happen that there exist some project-specific reasons which require certain ideally other type annotation members to be designed as String types. However, we do not find any sign of this during our check of the definitions of the involved annotations and we plan to contact some actual developers for confirmation of this in future. In addition, note even if project-specific reasons can potentially demand that certain annotation members must be of String type, this is bad practice that can incur high annotation maintenance costs as shown in Section 5.2.4. 
\begin{table}
 \caption{Results of Manually Checking Some String Type Annotation Members.}
\label{tab:manualchecking}
  \centering
  \small
  \begin{tabular}{|c|c|c|c|c|}
    \hline
  Matching Type & \#Total & \#Correct & \#Wrong & \#Unclear\\
    \hline
    Class Name & 100 & 34 & \color{blue}49 (Class)& 17\\
    Primitive & 100  & 66 & \color{blue}23 (Prim) & 11\\
    Annotation & 23 & 23 & - & - \\
    \hline
\end{tabular}
\end{table}

\begin{figure}[t]
\begin{lstlisting} 
package javax.jdo.annotations;
public @interface Column {
   ...
/*Whether null values are allowed to be inserted.*/
   String allowsNull() default ""; 
   ~(The annotation member should use Boolean type)~
}
(a) String Member that Should Use Primitive Type. 
 
package javax.annotation.sql;
public @interface DataSourceDefinition {
   ...
/*DataSource implementation class name.*/
   String className(); 
   ~(The annotation member should use Class type)~
}
  (b) String Member that Should Use Class Type. 
\end{lstlisting}
\caption{Examples of Wrong String Type Annotation Members.}
\label{lst:wrongdefinitionexample}
\end{figure}

\vspace{1.0mm} 
\noindent
\textbf{Finding 3}: \textit{Nearly three-quarter (73.71\%) of the assigned annotation values are typed as String, and the String content could actually be better typed (for instance, as Class or Primitive) in some cases. One important reason is that some annotation members are wrongly designed as String type.} 

\vspace{0.6mm}
\noindent
\textbf{Implication 3}: \textit{Developers could better benefit from static type checking with better annotation design. This calls for better support in IDEs to identify bad smell in annotation design and to suggest appropriate refactorings.} 


\section{Results on Annotation Evolution}
\label{sec:annotationevolution}
Like standard language constructs, annotations are in constant evolution. This section investigates the characteristics of annotation evolution.
\subsection{Overview of Annotation Evolution}
\label{sec:overallevolution}
We first want to see whether annotations are actively maintained by developers. To achieve this, we compare the churn rate of annotations with that of the entire code. The code churn rate is measured as \textit{Churned LOC}/\textit{Total LOC} \cite{icse05codechurn}, where \textit{Churned LOC} includes code lines added, modified or deleted. Similarly, we define churn rate for annotation as \textit{Churned annotation }/\textit{Total annotation}, where \textit{Churned annotation} includes 
the six different change types (defined in Section 3.2) made to annotations. We analyze each revision in the studied three year’s commit history to measure the churned code and annotations, and then calculate the churn rate for each year. The average churn rate of annotations is 0.95, 0.86, and 0.48 respectively for the considered three years, while the corresponding churn rate for the entire code is 0.92, 0.80, and 0.44 respectively. In summary, the churn rate of annotations is a little higher than that of the entire code, which implies that annotations are actively maintained by developers.

We then want to systematically investigate the overall evolution characteristics of annotations. \autoref{tab:annotationusechange} gives a summary of the six different change types (defined in Section 3.2) made to annotations during the considered three years of commit history. In this table, the \emph{Prevalence} and \emph{Percentage} columns give the actual number and percentage for each change type respectively, and the \emph{\#Commits} column shows the number of unique commits that each change type is involved. From the table, we can see that the two most frequent change types are "ADD\_ALL" and "DEL\_ALL", which in total have a percentage of 86.57\%. These two change types are obviously related with changes to other normal source code. For change type "ADD\_ALL", it reflects more about the developer's motivation of using annotations in the program. Compared with the relatively small percentage of change type "ADD\_ANN", the large percentage of change type "ADD\_ALL" implies that developers tend to decide whether to use annotations directly when they implement the code. For change type "DEL\_ALL", it is the by-product of deleting the annotated program elements. Overall, the significantly high percentage of change types "ADD\_ALL" and "DEL\_ALL" suggests that majority of annotation changes occur in parallel with changes to the annotated program elements, and they belong to \textit{code consistent annotation changes}, that is, annotation changes that are related with changes to other code. 

\begin{table}
 \caption{Summary of Changes Made to Annotations.}
  \label{tab:annotationusechange}
 \centering
  \footnotesize
  \begin{tabular}{|l|r|r|r|r|}
    \hline
    Change Type & Prevalence & Percentage & \#Commits \\
    \hline
    ADD\_ALL & 4,606,777 & 61.82\% &314,226 \\
    ADD\_ANN & 496,178 & 6.66\% & 78,738 \\
    DEL\_ALL & 1,844,793 & 24.75\% & 136,018 \\
    DEL\_ANN & 174,189  & 2.34\% & 47,079 \\
    CHANGE & 194,407 & 2.61\% & 12,945 \\
    UPDATE & 134,801 & 1.82\% & 27,872 \\
    \hline
\end{tabular}
\end{table}

Compared with \textit{code consistent annotation changes}, we call annotation changes that are not related with other code changes \textit{code independent annotation changes} and they are our focus in this paper. Besides change types "ADD\_ALL" and "DEL\_ALL", the other 4 change types "ADD\_ANN", "DEL\_ANN ", "CHANGE", and "UPDATE" are likely to reflect more directly developers' concerns over the annotations (missing annotations or problems with existing annotations), and thus are likely to be \textit{code independent annotation changes}. These 4 change types account for a non-neglectable 13.43\% (999,575 instances in total) of all the changes made to annotations, and we view this percentage as the upper bound of \textit{code independent annotation changes}. 

According to our method of calculating the 4 change types, there are some factors that can possibly make the calculated annotation changes \textit{code consistent annotation changes}. First, 
among the 999,575 change instances for the 4 change types, the annotated program elements are updated or moved for 161,594 (16.16\%) change instances and these instances can possibly be update changes with the annotated program elements. Second, even if the annotated program elements do not update or move, the annotation changes may be related to other code changes in the same or different Java class files in the same commit. To establish the lower bound of \textit{code independent annotation changes}, we further extract annotation changes in the following way. First, we center on annotation changes associated with classes that only involve changes to annotations, i.e., no changes to other code. As changes to annotations can incur changes to import statements, changes to import statements are not calculated as changes to other code. For change types "ADD\_ANN", "DEL\_ANN ", "CHANGE", and "UPDATE", there are 272,964, 121,379, 117,293, and 75,273 annotation change instances respectively for which an associated class involves changes to other code (the total number is 586,909). After this step, we have 412,666 (999,575--586,909) remaining annotation changes. Second, we further exclude 10,368 annotation changes for which the  class associated with the annotation change uses another class that is changed in the same commit. For example, the associated class inherits another changed class in the same commit or uses the changed class as an annotation value. Finally, we have 402,298 change instances left, which we believe should be \textit{code independent annotation changes} in general and account for 5.39\% of all the changes made to annotations. We view this percentage as the lower bound of \textit{code independent annotation changes}. 

Overall, the existence of a non-neglectable percentage (5.39\%-13.43\%) of \textit{code independent annotation changes} suggests that the current developer practice of using annotations in the code is not good enough and some developers tend to use annotations in a subjective and arbitrary way, 
introducing problems related with annotations (missing annotations or issues with existing annotations). After detecting the annotation problems, developers take efforts to address them as after-thoughts. 
This shows the importance of systematic testing of annotation behaviours and dedicated tool that can help the systematic testing process would be extremely helpful. 

\vspace{1.0mm} 
\noindent
\textbf{Finding 4}: \textit{Annotations are actively maintained. Most annotation changes are consistent changes together with other code changes, but there also exists a non-neglectable percentage (5.39\%-13.43\%) of annotation changes that are independent of other code changes.} 

\vspace{0.6mm}
\noindent
\textbf{Implication 4}: \textit{The current practice of using annotations is not good enough and some developers tend to use annotations in a subjective and arbitrary way, introducing annotation problems. Developers take efforts to address the annotation problems as after-thoughts, calling for dedicated tool that can support the process of systematic annotation testing.}

\subsection{Details of Code Independent Annotation Changes} 
\label{sec:annotationevolutionindependent}
To further study the characteristics of \textit{code independent annotation changes}, we sample some change instances for the 4 change types from the extracted 402,298 change instances and study them in detail. Taking the prevalence of each change type into account, to achieve 95\% confidence level and ±5\% confidence interval, we need to sample 
384, 381, 382, and 382 change instances for change types “ADD\_ANN”, "DEL\_ANN”, “CHANGE”, and “UPDATE" respectively. As the required sample size for each change type is nearly the same, we
conservatively sample 384 (the maximum value of the four numbers) change instances  for each change type. Note the sample with a larger sample size can also guarantee the targeted 95\% confidence level and ±5\% confidence interval.

For change types "ADD\_ANN" and "DEL\_ANN", we focus on studying the reason behind the annotation change behaviors. To achieve this, we collectively study the usage of the involved annotations, the annotated program elements, and the commit message. Whenever we are not clear about the reasons, we classify them to "Unknown". We remark that for the majority of the annotations involved in these two change types, we are familiar with their usage and the threats to validity are minor. 

\subsubsection{Annotation Addition}
For change type "ADD\_ANN", we observe that there are three major reasons behind the change behaviors for the sampled 384 change instances and \autoref{tab:reasonadd} gives a summary of the reasons. 

\begin{table}
 \caption{Summary of Reasons for Annotation Addition.}
  \label{tab:reasonadd}
  \centering
  \footnotesize
  \begin{tabular}{|p{4.5cm}|p{2.7cm}|}
  \hline
  Reason for Annotation Addition & \#Number(Percentage) \\
  \hline
    \hline
    Clear IDE Warning & 121 (31.5\%) \\
    \hline
    Clarify Specific Facts About Usage & 129 (33.6\%)\\
    \hline
    Introduce New Feature & 98 (25.5\%)\\ 
    \hline
    Unknown & 36 (9.4\%)\\ 
    \hline
\end{tabular}
\end{table}

\textbf{Clear IDE warning.} Among the 384 sampled instances for change type "ADD\_ANN", 31.5\% (121) of them are clearing the warnings raised by the IDE. In particular, there are 61, 49, 7, and 4 change instances for the annotations @\texttt{SuppressWarnings}, @\texttt{Override}, @\texttt{SuppressLint}, and @\texttt{SafeVarargs} respectively. 

\textbf{Clarify specific facts about usage.} In addition, 33.6\% (129) of the sampled change instances for change type "ADD\_ANN" are clarifying specific facts about usage of the annotated elements. These facts are majorly concerned with the scopes and conditions for the usages of the annotated elements, and \autoref{tab:annotationfacts} gives a summary of the involved annotations. We can see from the table that the most frequently involved annotation is @\texttt{Deprecated}, which specifies that the annotated program element is discouraged from using (either because it is dangerous or because there exists a better alternative). By clarifying specific facts about usage of the annotated elements through annotations, there should be less confusions and the code quality can possible be improved.

\begin{table}
 \caption{Annotations That Clarify Specific Facts About Usage.}
  \label{tab:annotationfacts}
  \centering
  \footnotesize
  \begin{tabular}{|l|c||l|c|}
    \hline
      Annotation & \#Number &  Annotation & \#Number \\
    \hline
    @\texttt{Deprecated} & 80 & @\texttt{Incubating} & 2\\
    @\texttt{Beta} & 9 & @\texttt{Unstable} & 2 \\
    @\texttt{ThreadSafe}  & 5 & @\texttt{SystemLevel} & 2 \\
    @\texttt{Internal} & 5  & @\texttt{Private} & 2\\
    @\texttt{NotThreadSafe} & 4 & @\texttt{TestApi} & 2 \\
    @\texttt{Experimental} & 3 & Others & 13 \\
    \hline
\end{tabular}
\end{table}

\textbf{Introduce new feature.} Lastly, 25.5\% (98) of the sampled instances are explicitly introducing new features. These features are concerned with the advanced annotation usages: static code analysis (58 change instances), run-time check (18 change instances), mapping and data binding (with HTTP request, XML file, JSON files, command line parameter, and databases, etc., 12 change instances), dependency injection (6 change instances), and code generation (4 change instances). The most frequently involved new code feature is static code analysis, due to the widely used and needed null-pointer analysis and the many nullability related annotations provided by different API providers. Again, these introduced advanced annotations usages could potentially lead to improved code quality and programmer productivity. 

In the analyzed samples, we also observe several change behaviours which move the annotation from one program element to another related program element. According to our method of calculating the six different change types defined in Section 3.2, this kind of "move" change will be calculated as one "add" change and one "delete" change. For the sampled 384 change instances for change type "ADD\_ANN", 9 are actually annotation move changes. Similarly, 13 of the 384 sampled change instances for change type "DEL\_ANN" are actually annotation move changes. \autoref{lst:movechange} gives an example of annotation move change, which moves the @\texttt{SuppressWarnings} annotation from method to class scope. Among the 22 move changes, three most frequently involved annotations are @\texttt{SuppressWarnings} (7 instances), @\texttt{Produces} (3 instances), and @\texttt{ConditionalOnClass} (2 instances). Moreover, \autoref{tab:movechange} shows the number of different kinds of move changes. We can see that the most frequently involved change is moving an annotation from method declaration to class declaration (8 instances), but meanwhile the second most frequently involved change is oppositely moving an annotation from class declaration to method declaration (5 instances). Overall, there are 14 move changes that move an annotation from the nested program element to the outer program element and 8 move changes that instead move an annotation from the outer program element to the nested program element. According to the official documentation by oracle for @\texttt{SuppressWarnings} \cite{annotationplace}, the general style is using an annotation on the most deeply nested element where it is effective. Our results suggest that some developers do not follow this style. The possible reason can be for code conciseness. For instance, a single @\texttt{SuppressWarnings} annotation on the class declaration can replace many @\texttt{SuppressWarnings} annotations on program elements of the class. 

\begin{figure}[t]
\begin{lstlisting} 
~+@SuppressWarnings("unused")~
public final class ACRAConfiguration {
   ~-@SuppressWarnings("unused")~
   public Map<String, String> getHttpHeaders() {
      
\end{lstlisting}
\caption{A Real Example of Annotation Move Change (Commit e79aac5d).}
\label{lst:movechange}
\end{figure}

\begin{table}
 \caption{Different Kinds of Annotation Move Changes}
  \label{tab:movechange}
  \centering
  \footnotesize
  \begin{tabular}{|l|c||l|c|}
    \hline
      Move Type & \#Number &  Move Type & \#Number \\
    \hline
    method$\rightarrow$class & 8 & field$\rightarrow$class & 2\\
    class$\rightarrow$method & 5 & class$\rightarrow$inner class & 1 \\
    method$\rightarrow$parameter & 2 & method$\rightarrow$package & 1 \\
    variable$\rightarrow$method & 2  & inner class$\rightarrow$class & 1\\
    \hline
\end{tabular}
\end{table}

\vspace{1.0mm} 
\noindent
\textbf{Finding 5}: \textit{For annotation add changes, three major reasons are clearing IDE warning (31.5\% ), clarifying specific facts about usage (33.6\%), and introducing new feature (25.5\%). There also exist some annotation move changes, which inconsistently move between outer and nested program elements.} 

\vspace{0.6mm}
\noindent
\textbf{Implication 5}: \textit{As after-thoughts, developers actively add annotations to the existing program elements to improve code readability and quality. When the annotations are effective, there sometimes exists a confusion about whether to put an annotation on the outer program element or the nested program element.} 

\subsubsection{Annotation Deletion}
For change type "DEL\_ANN", there are also three major reasons which cause the annotation deletion behaviors for the 384 sampled instances and \autoref{tab:reasondelete} gives a summary of the reasons. 

\begin{table}
 \caption{Summary of Reasons for Annotation Deletion.}
  \label{tab:reasondelete}
  \centering
  \footnotesize
  \begin{tabular}{|p{4.5cm}|p{2.7cm}|}
  \hline
  Reason for Annotation Deletion & \#Number(Percentage) \\
  \hline
    \hline
    Inconsistent Annotations & 121 (31.5\%) \\
    \hline
    Redundant Annotations. & 59 (15.4\%)\\
    \hline
    Wrong Annotations & 93 (24.5\%)\\ 
    \hline
    Unknown & 111 (28.6\%)\\ 
    \hline
\end{tabular}
\end{table}

\textbf{Inconsistent annotations.} Among the 384 sampled instances for the change type "DEL\_ANN", 31.5\% (121) of them delete the annotation as the annotated code was evolved but the annotation was not updated accordingly. Notably, there are two kinds of annotations that can easily become inconsistent but developers often forget to update them at first. First, annotations that suppress warning information. For instance, suppose a method $m_{1}$ previously calls another deprecated method $m_{2}$ and $m_{1}$ is annotated with @\texttt{SuppressWarnings("deprecation")}. However, $m_{2}$ is no longer deprecated in the new code version but the annotation on $m_{1}$ is still there. We have observed 93 change instances that are related with annotations for suppressing warnings in our sample (including both @\texttt{SuppressWarnings} and @\texttt{SuppressLint}), and these change instances typically are motivated by IDE warnings. Second, annotations that clarify specific facts about usage. For example, developers sometimes use annotation @\texttt{Incubating} to 
indicate that the code feature is currently a work-in-progress and may change at any time. However, they may forget to remove the annotation even if the code feature has been improved and finalized. There are 22 change instances that are related with annotations for clarifying specific facts about usage in our sample. 

\textbf{Redundant annotations.} Besides, 15.4\% (59) of the sampled 384 change instances delete the annotation as the annotation is redundant either at the beginning or because of other code update. In our sampled instances, 13 annotations are deleted as they become redundant as other code update and 46 annotations are deleted as they are redundant at the beginning. \autoref{lst:inconsistentchange} gives examples of these two kinds of redundant annotation deletion behaviors. With regard to deleting the redundant annotation because of other code change in \autoref{lst:inconsistentchange} (a), the original commit message is \textit{"no longer need to inject references since we are configuration based"}. In other words, the program uses annotation based dependency injection to inject the services at first and @\texttt{Reference} is used. Later on, the program changes to configuration file based injection, making the annotation @\texttt{Reference} redundant.

For deleting the annotation that is redundant at first in \autoref{lst:inconsistentchange} (b), there is no need to do null analysis on primitive type (boolean in the example) so that the @\texttt{NotNull} annotation is redundant. In our samples, we observe that there are many similar annotation uses that can easily be identified as unnecessary according to the characteristics of the annotated program element and the purpose of the annotation. In particular, we observe 1) nullability related annotations on primitive types (4 change instances); 2) nullability related annotations on methods that already have null check (2 change instances); 3) nullability related annotations on methods that return void (1 change instance); 4) @\texttt{Singleton} (identify a type that the injector only instantiates once) annotation on abstract class (2 change instances); 5) @\texttt{Singleton} annotation on static class (1 change instance); and 6) @\texttt{Inject} annotation on constructors for abstract classes (3 change instances, redundant as abstract classes cannot be instantiated). For those widely used annotations, tools can establish their unnecessary usage scenarios as redundant annotation templates and use the templates to detect redundant annotations in the code. 
 
\begin{figure}[t]
\begin{lstlisting} 
~-@Reference(policy = ReferencePolicy.DYNAMIC)~
protected void addOpenConnectIdConnectProvider {
   (a) Redundant Due to Code Update (Commit 67343bd).
~-@NotNull~
public boolean getIsContainer()
   (b) Redundant at the Beginning (Commit c392d26).

\end{lstlisting}
\caption{Real Examples of Deleting Redundant Annotations.}
\label{lst:inconsistentchange}
\end{figure}

\textbf{Wrong annotations.} In addition, 24.5\% (93) of the sampled 384 change instances delete the annotation as the annotation is wrong at the beginning. In particular, we observe the following 4 categories of annotations that are frequently used wrongly: annotations related with nullability analysis (27 change instances), annotations related with dependency injection (18 change instances), annotation related with mapping and data binding (9 change instances), and annotations related with concurrency (3 change instances). \autoref{lst:wrongannotations} gives examples of these kinds of wrongly used annotations. The original commit messages for the annotation deletions in \autoref{lst:wrongannotations} (a), (b), (c), and (d) are \textit{"getParent() is a containing directory which is nullable"}, \textit{"AddContactDialogView should not be a singleton"}, \textit{"The constructor for an abstract class is not really a factory"}, and \textit{"Fix Immutable annotation on TrackFileResource"} respectively. Overall, the large percentage of wrongly used annotations suggests the importance of testing annotations, in particular testing annotations related with nullability analysis, dependency injection, mapping and data binding, and concurrency. 

\begin{figure}[t]
\begin{lstlisting} 
~-@NotNull~
PsiElement getParent();
      (a) Nullability Analysis (Commit 0adf1d2).
~-@Singleton~
public class AddContactDialogView {
      (b) Dependency Injection (Commit 237fdc6).
~-@JsonCreator~
protected FileStrategyConfiguration (
        final FileStrategyConfiguration type
    (c) Mapping and Data Binding (Commit 9f9cb87).     
~-@Immutable~
public abstract class IMFTrackFileResourceType {
      (d) Concurrency (Commit 162a66e).
      
\end{lstlisting}
\caption{Real Examples of Deleting Wrong Annotations.}
\label{lst:wrongannotations}
\end{figure}

\vspace{1.0mm} 
\noindent
\textbf{Finding 6}: \textit{For annotation deletion changes, three major reasons are inconsistent (31.5\%), redundant (15.4\%), and wrong (24.5\%) annotations. Some redundant annotations can (easily) be identified by checking the attribute of the annotated program element. Most wrong annotations are related with nullability analysis, dependency injection, mapping and data binding, and concurrency.} 

\vspace{0.6mm}
\noindent
\textbf{Implication 6}: \textit{Tools can do static analysis to detect annotation bad smells: for instance some redundant annotations can be found with specific templates about unnecessary usage scenarios. Annotation testing is important, in particular for those kinds of annotations that can easily get wrong according to our finding.} 

\vspace{1.2mm}
For change types "CHANGE" and "UPDATE", the annotations exist both before and after the changes have been made and thus contain more specific facts about how the annotations themselves are changed. As a result, we focus on studying the detailed change behaviours for these two change types. Whenever we are not clear about
the change behaviors, we classify them to "Unknown". 

\subsubsection{Annotation Replacement}
With regard to change type "CHANGE", we observe two common annotation replacement behaviors in the sampled 384 instances. 
\autoref{tab:replacement} gives a summary of the replacement behaviors and \autoref{lst:changeexample} gives examples of them. 

\begin{table}
 \caption{Summary of Annotation Replacement Behaviors.}
  \label{tab:replacement}
  \centering
  \footnotesize
  \begin{tabular}{|p{4.5cm}|p{2.7cm}|}
  \hline
  Replacement Behavior & \#Number(Percentage) \\
  \hline
    \hline
    Switch to the Opposite Annotation & 124 (32.3\%) \\
    \hline
    Switch to the "Same-name" Annotation in Another Library & 76 (19.8\%)\\
    \hline
    Unknown & 184 (47.9\%)\\ 
    \hline
\end{tabular}
\end{table}

\begin{figure}[t]
\begin{lstlisting} 
-protected void innerKill(~@Nullable~ Throwable t) {
+protected void innerKill(~@Nonnull~ Throwable t) {
}
(a)Switch to Opposite Annotation (Commit 23c777e).

public class CreateInvoiceNumerators {
  ~-@com.google.inject.Inject~
  ~+@javax.inject.Inject~
  PropertyRepository propertyRepository;
(b)Switch to "Same-name" Annotation (Commit 53b2170).

\end{lstlisting}
\caption{Real Examples of Annotation Replacements.}
\label{lst:changeexample}
\end{figure}

\textbf{Switch to the opposite annotation.} Among the 384 change instances for change type "CHANGE", 32.3\% (124) of them are replacing one annotation with another annotation that gives opposite metadata. \autoref{tab:opposite} shows the observed different kinds of opposite annotation changes and the corresponding frequency. Switching an annotation to its opposite annotation typically implies an ``annotation bug'', i.e., the information encoded in the annotation metadata is wrong. Compared with other kinds of annotation replacement changes, opposite annotation replacement changes are more tractable and the large percentage of this kind of annotation change suggests an interesting avenue for future research: automatically detecting and repairing annotation bugs that require replacing one annotation with its opposite annotation. 

\begin{table}
 \caption{Summary of Opposite Annotation Changes}
  \label{tab:opposite}
  \centering
  \footnotesize
  \begin{tabular}{|l|r|r|r|r|r|}
    \hline
      Opposite Annotation Changes & Frequency \\
    \hline
    @\texttt{Nullable}$\leftrightarrow$@\texttt{Notnull} & 80 \\
    @\texttt{Private}$\leftrightarrow$ @\texttt{Public} & 12 \\
    @\texttt{GET}$\leftrightarrow$ @\texttt{POST} & 8 \\
    @\texttt{*Many*}$\leftrightarrow$ @\texttt{*One*} & 6 \\
    @\texttt{ThreadSafe}$\leftrightarrow$ @\texttt{NotThreadSafe} & 4 \\
     @\texttt{Stable}$\leftrightarrow$ @\texttt{Unstable} & 4 \\
 @\texttt{GwtIncompatible}$\leftrightarrow$ @\texttt{GwtCompatible} & 2\\
    \texttt{Others} & 8\\
    \hline
\end{tabular}
\end{table}

\textbf{Switch to the "same-name" annotation in another library.} In addition, 19.8\% (76) of the 384 change instances for change type "CHANGE" are replacing one annotation with another annotation that has the same name but comes from a different library (i.e., a different API provider). In particular, for the sampled change instances, we have observed that this kind of annotation change happens frequently for nullability related annotations (26 instances, e.g.,@\texttt{Nullable}), serialization or persistence related annotations (9 instances, e.g., @\texttt{Column}), and dependency injection related annotations (8 instances, e.g.,@\texttt{Inject}). For annotations involved in this kind of change, there are typically several API providers that provide different implementations for the same annotations. For instance, jetbrains, android, eclipse, and checkerframework \cite{type-checker} all have their respective implementations for the nullability related annotation @\texttt{Nullable}. The large number of annotation changes in this category suggests that developers are often confused about which specific annotation to use when several API providers all provide an implementation of the annotation, calling for a unified standardization among the community. 

\vspace{1.0mm} 
\noindent
\textbf{Finding 7}: \textit{For annotation replacement changes, 32.3\% of them are switching to opposite annotations and 19.79\% of them are switching to "same name" annotations from other libraries.} 

\vspace{0.6mm}
\noindent
\textbf{Implication 7}: \textit{There exists the prospect of automatically detecting and repairing annotation bugs that use opposite annotations of the correct ones, and unified standardization for popular annotations with different API providers will be beneficial.}

\subsubsection{Annotation Value Update}
\label{sec:annotationupdate}
For change type "UPDATE", we observe 4 different kinds of annotation value update behaviors in the sampled 384 instances. These annotation value update behaviors are summarized in \autoref{tab:updatevalue} and exemplified in \autoref{lst:updateexample}. 

\begin{table}
 \caption{Summary of Annotation Value Update Behaviors.}
  \label{tab:updatevalue}
  \centering
  \footnotesize
  \begin{tabular}{|p{4.5cm}|p{2.7cm}|}
  \hline
  Update Behavior & \#Number(Percentage) \\
  \hline
    \hline
    Correct String of Class Name & 40 (10.4\%) \\
    \hline
    Correct String of Code & 45 (11.7\%)\\
    \hline
    Improve String of Static Text & 92 (23.9\%) \\
    \hline
    Override or Restore Default Value & 109 (28.4\%) \\
    \hline
    Unknown & 98 (25.6\%)\\ 
    \hline
\end{tabular}
\end{table}

\begin{figure}[t]
\begin{lstlisting} 
~-@ConditionalOnBean~!(type="org.apache.camel.springboot.CamelAutoConfiguration")!
~+@ConditionalOnBean~!(type="org.apache.camel.spring.boot.CamelAutoConfiguration")!
public class AhcComponentAutoConfiguration {
   (a)Correct String of Class Name (Commit 9c01dc0)
   
~-@Specialization~!(guards="!isObject")!
~+@Specialization~!(guards="!isObject(frame, vector)")!
protected RStringVector doStringVector {
     (b)Correct String of Code (Commit 97b4f49)
     
~-@Description~!("If dbms.killQueries give a verbose output, with information about which querys where not found.")!
~+@Description~!("Specifies whether or not dbms.killQueries produces a verbose output, with information about which queries were not found")!
public static Setting<Boolean> kill_query_verbose =
  (c)Improve String of Static Text (Commit 5ff39df)
  
~-@DatabaseField~!(unique = true)!
~+@DatabaseField~!(id = true,unique = true)!
protected String code;
(d)Override or Restore Default Value(Commit 33b9ec5)
\end{lstlisting}
\caption{Real Examples of Annotation Value Updates.}
\label{lst:updateexample}
\end{figure}

\textbf{Correct String of Class Name.} Among the 384 change instances for change type "UPDATE", 10.4\% (40) of them are correcting the class name represented as String content. The reason behind this kind of update behavior is described in Section 4.3. For the annotation members involved in this kind of update behaviour, it is probably better to design their types as Class and then compilers can type check the class name if developers make typo error or forget to update the class name after code refactoring. Overall, the large percentage of update behaviors in this category further confirms our concern about wrongly designing some annotation members as String type (in our samples, we also observe one change instance which corrects "tyre" with "true" due to wrongly design a boolean type annotation member as String type), and shows that this problem is particularly serious for some annotation members that ideally should be designed as Class type. 

\textbf{Correct String of code.} Another 11.7\% (45) of the 384 change instances for change type "UPDATE" are correcting some source code inside the String. Among the 45 instances in this category, 20 are correcting SQL statement. 13 are correcting regular expression, and 12 are correcting other code. The large percentage of code related update behaviors suggests the importance of testing String type annotation members that contain code as the content. Meanwhile, plenty of these update behaviors are syntax related (e.g., variable or function name change, keyword missing, code structure imbalance, typo, etc). 
Advanced IDE could try to recognize these code related annotation members (e.g., by checking the structure of String content and searching for code related keyword) and do syntax check to highlight potential problems about the code. 

\textbf{Override or Restore Default Value.} In addition, 28.4\% (109) of the 384 change instances for change type "UPDATE" are overriding or restoring the default values for some annotation members. Among the 109 instances, 84 instances are overriding the default values (i.e., use new values other than the default values specified in the annotation type declarations) and 25 instances are restoring the default values (i.e., discard the newly specified values and return to the default values specified in the annotation type declaration). In particular, 69.7\% (76 instances) of the 109 instances are concerned with switching between boolean value true and boolean value false for a certain annotation member. The large percentage of annotation value update behaviors in this category suggests that developer can easily make mistakes when dealing with annotation members with default values (especially boolean type annotation members). In particular, they are inclined to use default values and forget to adjust them according to the actual need. Overall, this problem highlights the importance of well testing and documenting annotation members with default values. 

\textbf{Improve String of Static Text.} Finally, 23.9\% (92) of the 384 change instances for change type "UPDATE" are improving static text. The static text typically is for the purpose of comment or documentation. Developers change the text typically for three reasons. First, the meaning of the text is not clear enough and developers improve the text for sake of clarification. Second, the text contains some spelling, grammar, or typo errors. Finally, the text is out of date. For instance, the text may contain some variable or function names that have been changed. 

\vspace{1.0mm} 
\noindent
\textbf{Finding 8}: \textit{For annotation value update changes, four typical update behaviors are overriding or restoring default value (28.4\%), improving string of static text (23.9\%), correcting string of code (11.7\%), and correcting string of class name (10.4\%).} 

\vspace{0.6mm}
\noindent
\textbf{Implication 8}: \textit{Testing annotations should pay special attention to annotation members with default values (especially boolean type members) and String type annotation members that contain code as the content. Advanced IDE could also highlight potential syntax problems with code used as (String type) annotation value.} 

\vspace{1.6mm}
Besides, we also make an observation about the change manner of \textit{code independent annotation changes}.
\subsubsection{Change Manner--Single Change vs Group Change} We use "Single Change" to refer to the scenario where a specific annotation change (e.g., replacing @\texttt{Nullable} with @\texttt{Notnull}) appears exactly once in a single commit, and we use "Group Change" to refer to the scenario where a specific annotation change appears multiple times in a single commit. We count how many times the "Single Change" and "Group Change" have happened for the sampled instances, and the results are given in \autoref{tab:singleorgroup}. For "Group Change", we also calculate the average number of times that a specific annotation change appears in a single commit, and the resultant number is given in the column \emph{Average Times}. 

\begin{table}
  \caption{Occurrences of "Single Change" and "Group Change" for the Sampled Change Instances.}
  \centering
  \footnotesize
  \label{tab:singleorgroup}
  \begin{tabular}{|l|c|c|c|r|}
    \hline
    \multirow{2}{*}{{Change Type}} & \multicolumn{1}{c|}{Single Change} & \multicolumn{2}{c|}{Group Change} \\ 
    \cline{2-4}
     & \#Instance  & \#Instance  & \#Average Times \\
    \hline
    \hline
ADD\_ANN & 31 (8.07\%) & 353 (91.93\%)  & 10.18 \\
DEL\_ANN & 79 (20.57\%) & 305 (79.43\%)  & 9.95 \\
CHANGE & 89 (23.17\%) & 295 (76.83\%) & 25.12 \\
UPDATE & 81 (21.09\%) & 303 (78.91\%) & 16.89 \\
    \hline
    \hline
Total & 280 (18.22\%) & 1,256 (81.78\%)  & 15.25\\
    \hline
  \end{tabular}
\end{table}

It can be seen from \autoref{tab:singleorgroup} that for any of the 4 change types, most annotation changes are "Group Changes". Across all the 4 change types, 81.78\% of the annotation changes are "Group Changes". Moreover, the average number of times that a specific annotation change appears in a single commit is also high, and it is 15.25 across all the 4 change types. For a certain annotation that is changed in "Group Change" manner in a commit, the different annotated program elements for this annotation in general are related in some ways. For instance, the annotated different program elements frequently are different subclasses of a certain superclass or similar members of the different subclasses. For another example, it is also quite common that the annotated different program elements are different overloading methods in the same class or similar members of the overloading methods. This characteristic not only suggests that developers should pay attention to other related program elements when they make changes to annotations for a certain program element, but also shows the feasibility of annotation change recommendation tools. When developers have finished the annotation change for a certain program element, annotation change recommendation tools can use program analysis techniques to analyze the program element involved in the change and search for other related program elements which likely will subject to the same annotation change. For annotation change types "DEL\_ANN", "CHANGE", and "UPDATE", the accuracy can further be improved as these three change types involve changes to existing annotations and thus related program elements can further be confined to ones annotated with the existing annotations. This indeed is promising and valuable as we  observe that it is quite common that several adjacent commits have exactly same annotation changes that apply to related program elements, which implies that developers frequently forget to make annotation changes for other related program elements when they make annotation changes for certain program elements.

\vspace{1.0mm} 
\noindent
\textbf{Finding 9}: \textit{For code independent annotation changes, most of them (81.78\%) change in the "Group Change" manner, i.e., multiple exactly same annotation changes happen in a single commit.} 

\vspace{0.6mm}
\noindent
\textbf{Implication 9}: \textit{We envision annotation change recommendation tools to help developers with the annotation change process. For instance, by analyzing some manually finished changes or recent changes, they could recommend the remaining changes to be done.}


\section{Results on Annotation Impact}
\label{sec:annotationimpact}
The annotation metadata is helpful for software development and it is argued that the use of annotations can reduce the likelihood of errors in the code. The argument arises as typical annotation use cases are likely to make code less error-prone. For annotations processed before run-time, two typical use cases are for static analysis and boilerplate code generation. Static analysis annotations can raise code problems early \cite{type-checker,icse17type} and code generation annotations partially replace the error-prone process of implementing code \cite{tutoriallombok}. For annotations processed at run-time, they are typically used for replacing conventional "side files" maintained in parallel with programs, and maintaining information directly in the program can simplify the development process and thus make code less error-prone \cite{tutorialoracleannotation}. To shed some light on whether the argument is valid in practice, we use the regression model described in Section 3.3 to investigate the correlation between annotation usage and code error-proneness. 

For the considered three years of commit history, we have collected in total 1,296,342 data instances, i.e., files with data about response variable, experimental variable, and control measures described in Section 3.3. Among the data instances, the response variable \textbf{nbbug} is zero for 960,651 data instances, which means 74.1\% of the files do not have any bug-fixing commits during the commit history considered. The hurdle regression model, which involves hurdle model for zero counts and count model for positive counts, is an ideal model for dealing data instances with "excess zero" characteristic. For our setting, the hurdle model corresponds to asking "Is there an association between using annotations and there being or not a fault in the file's history?". With the hurdle overcome, the count model equates to asking "For those files with at least one fault in their history, is there an association between using annotations and the number of faults in their history?".

\subsection{Overall Impact} 
We first use the hurdle regression model on the entire 1,296,342 data instances, and the results are presented in \autoref{tab:wholedata}. The left column contains the coefficients of the hurdle model and the corresponding standard errors, and the right column contains the coefficients of the count model and the corresponding standard errors. It can be seen from the table that the coefficients for annotation uses are negative in both the hurdle and count models, \emph{which indicates that the use of annotations is negatively correlated with the likelihood of errors in the code}. We can also see that the coefficients for the three control measures (\textbf{codeside}, \textbf{nbCommit}, and \textbf{nbDev}) are all positive in both models, which means that they are positively related with defect occurrence and this result is in line with previous studies \cite{premregressionmodel,commitclassify3, commitclassify1, icse05codechurn} on the impact of these control measures on code error-proneness. Compared with the coefficients of the three control measures, the absolute values of the coefficients for annotation uses are much smaller in both models, which implies that the control measures have a dominant impact on code error-proneness. Note even though the absolute values of the coefficients for annotation uses are small, they are significant in both models (\textit{p} value is less than 0.001).

\begin{table}[!htbp] 
\footnotesize
\centering 
  \caption{Results of Hurdle Regression Model for the Whole Data.} 
  \label{tab:wholedata} 
\begin{tabular}{@{\extracolsep{5pt}}lll} 
\hline 
\hline
 & \multicolumn{2}{c}{\textit{Dependent variable - coef. (p-value)}} \\ 
\cline{2-3} 
 & \multicolumn{2}{c}{nbbug} \\ 
 & hurdle model & count model \\ 
\hline 
  (Intercept) & -3.222978$^{***}$(0.008254) & -1.83e+00$^{***}$(7.66e-03) \\ 
 log(codesize) & 0.206817$^{***}$(0.002158) &  4.51e-01$^{***}$(1.59e-03) \\ 
 nbCommit & 0.212222$^{***}$(0.000936) & 1.24e-02$^{***}$(3.92e-05) \\ 
 nbDev & 0.353419$^{***}$(0.002594) & 2.48e-02$^{***}$(3.30e-04) \\ 
 nbAnnotation &-0.004115$^{***}$ (0.000258) & -3.73e-05$^{***}$ (7.99e-05) \\ 
\hline 
\hline 
\textit{Note:} & \multicolumn{2}{r}{$^{*}$p$<$0.1; $^{**}$p$<$0.05; $^{***}$p$<$0.001} \\ 
\end{tabular} 
\end{table} 

To further justify the use of hurdle regression model in our scenario where data instances have "excess zero" characteristic, we also fit the poisson regression model to the entire data and compare the corresponding residual sum of squares (RSS) \cite{RSS} with that of the hurdle regression model. RSS is the sum of the squared differences between the actual data values and the predicted data values, and the smaller the RSS, the better the model fit. For our data, the RSS is 1,147,741 for the hurdle regression model and is 1,635,745 for the poisson regression model, which obviously shows that the hurdle regression model is the better model for our problem. 

\subsection{Impact on Different Groups} 
We further study the relation between annotation uses and code error-proneness when the control measures are big and small respectively. To this end, for each of the three control measures, we split the entire data instances into two groups according to the median value of the control measure, one group for which the control measure is less than or equal to the median value, and the other group for which the control measure is larger than the median value. We then apply the hurdle regression model to each group and compare the result of one group with that of the other group.

\subsubsection{Group with different developer numbers} 
For control measure \textbf{nbDev}, the median value of the entire data instances for it is 1. To make both groups have the same control measures, we split the entire data instances by developer number according to value 2 (that is, $nbDev \leq 2$ and $nbDev\textgreater2$), and the results for applying the hurdle regression model to both groups are shown in \autoref{tab:developer}.  It can be seen from the table that the coefficients for annotation uses in hurdle and count models are still negative in both groups, and the absolute values of the two coefficients in the "Fewer Developers" group is much larger than that in the "More Developers" group, which suggests that the relation between using annotations and code error-proneness is larger when fewer developers are involved with the code. This happens perhaps because annotations are typically used by a few contributors with high ownership of the code (see Section 4.2), and the "minor contributors" with low ownership will ignore them. For "More Developers" group, it is likely that there will be more less "minor contributors" with low ownership of the code.

\begin{table*}[!htbp] \centering 
  \caption{Comparison of Results of Hurdle Regression Model for Data with Different Number of Developers.} 
  \label{tab:developer} 
\begin{tabular}{@{\extracolsep{5pt}}lcccc} 
\hline 
\hline 
 & \multicolumn{4}{c}{\textit{Dependent variable:}} \\ 
\cline{2-5} 
 & \multicolumn{4}{c}{nbbug (\textbf{Fewer Developers}) \hspace{3.1cm}    nbbug (\textbf{More Developers})} \\ 
& hurdle model & count model & hurdle model & count model \\ 
\hline 
  (Intercept) & -3.9535598$^{***}$(0.01146) & -2.580e+00$^{***}$(0.01593) & -1.3107329$^{***}$(0.02184) & -1.046e+00$^{***}$(0.00978) \\ 
 log(codesize) & 0.2182401$^{***}$(0.00246) & 4.222e-01$^{***}$(0.00289) & 0.1491194$^{***}$(0.00485) & 3.587e-01$^{***}$(0.00199)  \\ 
 nbCommit & 0.2648405$^{***}$(0.00119) & 2.522e-02$^{***}$(0.00008) & 0.1254791$^{***}$(0.00136) & 1.214e-02$^{***}$(0.00004) \\ 
 nbDev & 0.7403687$^{***}$(0.00566) & 2.950e-01$^{***}$(0.00614) & 0.0525988$^{***}$(0.00456) & 1.101e-02$^{***}$(0.00038)  \\ 
 nbAnnotation & -0.0052055$^{***}$(0.00032) & -2.142e-03$^{***}$(0.00021) & -0.0005084(0.00044) & -3.876e-05 (0.00008) \\ 
\hline 
\hline 
\textit{Note:}  & \multicolumn{4}{r}{$^{*}$p$<$0.1; $^{**}$p$<$0.05; $^{***}$p$<$0.001} \\ 
\end{tabular} 
\end{table*} 

\subsubsection{Group with different commit numbers} 
For control measure \textbf{nbCommit}, the median value of the entire data instances for it is 2. We split the entire data instances by commit number according to this value (that is, $nbCommit \leq 2$ and $nbCommit\textgreater2$), and the results for applying hurdle regression model to both groups are shown in \autoref{tab:commit}. We again see from the table that the coefficients for annotation uses in hurdle and count models are negative in both groups, and the absolute values of the two coefficients in the "Fewer Commits" group is larger than that in the "More Commits" group, which suggests that the relation between using annotations and code error-proneness is larger when the code is less changed. The possible reason is with more commits made to the code, the annotations themselves are also more likely to subject certain kinds of changes defined in Section 3.2. In other words, the life-cycle of the annotations are relatively short or the annotations are more likely to have some problems, both of which will impact the value of annotations. 

\begin{table*}[!htbp] \centering 
  \caption{Comparison of Results of Hurdle Regression Model for Data with Different Number of Commits.} 
  \label{tab:commit} 
\begin{tabular}{@{\extracolsep{5pt}}lcccc} 
\hline 
\hline 
 & \multicolumn{4}{c}{\textit{Dependent variable:}} \\ 
\cline{2-5} 
 & \multicolumn{4}{c}{nbbug (\textbf{Fewer Commits}) \hspace{3.8cm}    nbbug (\textbf{More Commits})} \\ 
& hurdle model & count model & hurdle model & count model \\ 
\hline 
  (Intercept) & -4.9545597$^{***}$(0.01783) & -29.809700(33.55469)  & -2.1011866$^{***}$(0.01209) & -1.323e+00$^{***}$(0.00797) \\ 
 log(codesize) & 0.2119938$^{***}$(0.00328) & 0.306861$^{***}$(0.01221) & 0.2046953$^{***}$(0.00301) & 3.804e-01$^{***}$(0.00166) \\ 
 nbCommit & 1.0745945$^{***}$(0.00913) & 13.637954(16.77733)  & 0.1255522$^{***}$(0.00091) & 1.257e-02$^{***}$(0.00004) \\ 
 nbDev & 0.43682158$^{***}$(0.00941)  & -0.171596$^{***}$(0.02786) & 0.1886655$^{***}$(0.00247) & 1.823e-02$^{***}$(0.00034) \\ 
 nbAnnotation & -0.0055438$^{***}$(0.00051) & -0.013309$^{***}$(0.00221) & -0.0032436$^{***}$(0.00029) & -1.053e-04(0.00008) \\ 
\hline 
\hline 
\textit{Note:}  & \multicolumn{4}{r}{$^{*}$p$<$0.1; $^{**}$p$<$0.05; $^{***}$p$<$0.001} \\ 
\end{tabular} 
\end{table*}

\subsubsection{Group with different code size} 
For control measure \textbf{codesize}, the median value of the entire data instances for it is 32. Again, we split the entire data instances by code size according to this value (that is, $codesize \leq 32$ and $codesize\textgreater32$), and the results for applying the hurdle regression model to both groups are shown in \autoref{tab:filesize}. It can be seen from the table that the coefficients for annotation uses in both hurdle and count models become positive for "Smaller Size" group, and they are still negative for "Larger Size" group. This result suggests that the positive impact of annotations on making code less error-prone lies more in big files, and for small files, annotation uses can even have a negative impact. The possible reason is that small files are relatively simple in function and logic, which makes developers less likely to make mistakes. Consequently, the value of the additional information provided by annotations decreases for these small files. The negative impact for small files arises perhaps because developers can possibly use annotations in a wrong manner (see Section 5.2). 

\begin{table*}[!htbp] \centering 
  \caption{Comparison of Results of Hurdle Regression Model for Data with Different Code Size.} 
  \label{tab:filesize} 
\begin{tabular}{@{\extracolsep{5pt}}lcccc} 
\hline 
\hline 
 & \multicolumn{4}{c}{\textit{Dependent variable:}} \\ 
\cline{2-5} 
 & \multicolumn{4}{c}{nbbug (\textbf{Smaller Size}) \hspace{4.0cm}    nbbug (\textbf{Larger Size})} \\ 
& hurdle model & count model & hurdle model & count model \\ 
\hline 
 (Intercept) & -3.3744804$^{***}$(0.01478)& -2.6490524$^{***}$(0.03237) & -2.9282814$^{***}$(0.01954)  & -1.496e+00$^{***}$(0.01052) \\ 
 log(codesize) & 0.1327494$^{***}$(0.00551) & 0.6009414$^{***}$(0.01146)  & 0.1794751$^{***}$(0.00435) & 3.897e-01$^{***}$(0.00211) \\ 
 nbCommit & 0.2364088$^{***}$(0.00209) & 0.0132305$^{***}$(0.00028) & 0.2090518$^{***}$(0.00104) & 1.263e-02$^{***}$(0.00004) \\ 
 nbDev & 0.5033255$^{***}$(0.00479) & 0.0888968$^{***}$(0.00194) & 0.2797320$^{***}$(0.00303) & 2.363e-02$^{***}$(0.00034) \\ 
 nbAnnotation & 0.0006745(0.00179) & 0.0289408$^{***}$(0.00232) & -0.0037693$^{***}$(0.00026) & -4.219e-04$^{***}$(0.00008) \\ 
\hline 
\hline 
\textit{Note:}  & \multicolumn{4}{r}{$^{*}$p$<$0.1; $^{**}$p$<$0.05; $^{***}$p$<$0.001} \\ 
\end{tabular} 
\end{table*}

\vspace{1.0mm} 
\noindent
\textbf{Finding 10}: \textit{There exists a small but significant relationship between annotation uses and code error-proneness: Java code with annotations tends to be less error-prone. The relationship is larger for bigger files as well as for files with fewer developers and commits.}

\vspace{0.6mm}
\noindent
\textbf{Implication 10}: \textit{Java annotations in general should be encouraged to use as it can potentially lead to an improvement in software quality.}

\section{Threats to Validity}
\label{sec:threat}
\vspace{1.0mm} 
\textbf{Construct Validity.} We search error related key words in the commit message to identify bug-fixing commits and use them as a proxy for defect occurrences. It is possible that developers do not use or use some other error related key words to describe a bug-fixing commit. Meanwhile, the commit messages for some other kinds of commits can possibly have error related key words. Thus, our classification method can incorrectly classify some commits. However, due to the large size of our dataset, it is extremely difficult to manually check each commit message. We sample some commits and the manual analysis shows that the accuracy of the classification is acceptable. In addition, several previous studies \cite{icsmbug}\cite{commitclassify1}\cite{commitclassify2}\cite{commitclassify3}\cite{commitclassify4} have adopted the same method to classify commits.

\vspace{1.0mm} 
\textbf{Internal Validity.} 
We sample 1,536 change instances for the 4 change types "ADD\_ANN", "DEL\_ANN", "CHANGE", and "UPDATE", and manually check them to study  
the characteristics of code independent annotation changes. 
This process may introduce errors. To reduce this threat as much as possible, the reported results about characteristics of code independent annotation changes are checked
and confirmed by two authors of the paper. In addition, the
complete results are made publicly available online to let readers
gain a more deep understanding of our study and analysis.

\vspace{1.0mm} 
\textbf{External Validity.} We use projects hosted on Github in this study, a potential threat to validity is whether the results will generalize to other projects hosted on other platforms. To reduce this threat, we use a large number of projects and these projects cover various domains. In addition, the used projects are all open-source projects and it can be possible that the characteristics of annotation uses in closed-source projects are different. 

\section{Related work}
\label{sec:relatedwork}
\vspace{1.0mm} 
\textbf{Empirical Study on Java Annotations.} Using 106 open source Java projects as subjects, Rocha and Valente \cite{annotationseke} empirically investigate what are the most widely used annotations and what kinds of program elements are more likely to be annotated. Compared with their work, our study is much more large scale and targets more fundamental questions related with annotation usage, annotation evolution, and annotation impact. Parnin et al. \cite{genericsemse} use 40 open-source Java projects to study the adoption of Java generics and contrast it with the adoption of annotations. Dyer et al. \cite{annotationicse} use projects on SourceForge to find uses of 18 Java language features over time, including the annotation feature. Both these two studies have shown that annotations are mainly adopted by few people in the development team, but it is not clear what kinds of developers are more likely to use annotations. 

\vspace{1.0mm} 
\textbf{Java Annotation Related Techniques.} To detect more errors during compile time, Papi et al. \cite{isstatypechecker} introduce a pluggable type checking system called \textit{Checker Framework}. The framework first uses Java annotations to define type qualifiers and then uses annotation processing at compile time to enforce the semantics of the defined annotations. The type checkers can be used for achieving different kinds of checking, such as nullness analysis \cite{type-checker}, reference immutability\cite{oopslatype}, implicit control flow \cite{asetype}, locking discipline \cite{Ernst2016} and class immutability \cite{icse17type}. In light of the fact that a large amount of boilerplate code is used to traverse the Abstract Syntax Trees (ASTs), Zhang et al. \cite{OOPSLA15} present a framework called \textit{Shy} which makes use of compile time annotation processing to generate generic code for various types of traversals. 

\vspace{1.0mm} 
\textbf{Empirical Study on Other Java Language Features.} Basit et al. \cite{SEKEGenrics} use two case studies to investigate the usefulness of generics in removing redundant code. Tempero et al. \cite{Javaheritance} empirically investigate the use of inheritance and the inheritance structures in real-world programs. Tempero \cite{JavaField} also conducts an empirical study to see whether the principle of avoiding use of non-private fields is followed in practice. Grechanik et al. \cite{ESEMJava} use a large corpus to study how a set of language features are used by developers in practice. The studied features do not include new features introduced after JLS2 \cite{JLS2} such as annotation and generics. Hoppe and Hanenberg \cite{OOSPLAGenerics} empirically explore whether the use of Java generic can actually result in an increased developer productivity. To investigate how Java developers have adapted to the functional style of thinking, Mazinanian et al. \cite{oopsla2017} perform an empirical study to understand the use of Lambda Expressions in Java. Due to the increasing importance of optimizing energy consumption, Hasan et al. \cite{ICSE16energy} investigate how different common operations on Java collections classes will consume energy. Kochhar and Lo \cite{commitclassify4} investigate how assertion uses are related with defect occurrence, code ownership and developer experience. Dietrich et al. \cite{ecoopcontract} perform an empirical study to characterize the use and evolution of contracts in real Java programs. To better understand the usefulness and limitations of existing static analysis techniques in dealing with code that uses Java reflection API, Landman et al. \cite{icsereflection} conduct an empirical study to characterize the usage of Java reflection API. 

\vspace{1.0mm} 
\textbf{Annotation Languages.} Many annotation languages have been proposed for different kinds of programming languages. These annotation languages typically aim to conduct certain kinds of program analysis. For Java, even before annotation is introduced as a language feature, there exist annotation languages used for verification and debugging \cite{OOPSLA15}, for compile time checking \cite{ESCChecker}, and for both verification and compile time checking \cite{Alloy}. For C, annotation languages have been used for annotating control flows and function interface \cite{tutorialmsdn}\cite{cannotation}, for expressing synchronization assumptions \cite{Tanannotation}, and for specifying locking requirements \cite{spark}. For Ada, there exists an annotation language named ANNA \cite{anna} that can be used for specifying subprograms, packages, exceptions, and contexts.

\section{Conclusion}
\label{sec:conculsion}
Despite the ever-growing use of Java annotations, there is still limited empirical knowledge about the actual usage of annotations in practice, the changes made to annotations during software evolution, and the potential impact of annotations on code quality. In this paper, we perform the first large-scale empirical study about Java annotation usage, evolution, and impact over 1,094 open-source projects hosted on GitHub. Our study generates 10 interesting findings with important implications. These implications shed light for developers, researchers, tool builders, and language or library designers in order to improve all facets of Java annotation engineering. 

\section*{Acknowledgment}

We are grateful to the anonymous reviewers for their insightful comments. Zhongxing Yu and Martin Monperrus are partially supported  by the Wallenberg AI, Autonomous Systems and Software Program (WASP) funded by the Knut and Alice Wallenberg Foundation. Chenggang Bai is supported in part by the National Natural Science Foundation of China under Grant 61772055 and Equipment Preliminary R\&D Project of
China under Grant 41402020102. 

\bibliographystyle{abbrv}
\balance
\bibliography{references} 

\begin{thebibliography}{10}

\bibitem{spark}
L. torvalds. sparse-a semantic parser for c.
\newblock \url{https://www.kernel.org/pub/software/devel/sparse/}, 2018.

\bibitem{asetype}
P.~Barros, R.~Just, S.~Millstein, P.~Vines, W.~Dietl, M.~dAmorim, and M.~D.
  Ernst.
\newblock Static analysis of implicit control flow: Resolving java reflection
  and android intents (t).
\newblock In {\em Proceedings of the 2015 30th IEEE/ACM International
  Conference on Automated Software Engineering (ASE)}, ASE '15, pages 669--679,
  Washington, DC, USA, 2015. IEEE Computer Society.

\bibitem{SEKEGenrics}
H.~A. Basit, D.~C. Rajapakse, and S.~Jarzabek.
\newblock An empirical study on limits of clone unification using generics.
\newblock In {\em In Proceedings of the 17th International Conference on
  Software Engineering and Knowledge Engineering (SEKE’05}, pages 109--114,
  2005.

\bibitem{ownership2}
C.~Bird, N.~Nagappan, B.~Murphy, H.~Gall, and P.~Devanbu.
\newblock Don't touch my code!: Examining the effects of ownership on software
  quality.
\newblock In {\em Proceedings of the 19th ACM SIGSOFT Symposium and the 13th
  European Conference on Foundations of Software Engineering}, ESEC/FSE '11,
  pages 4--14, New York, NY, USA, 2011. ACM.

\bibitem{commitclassify1}
C.~Casalnuovo, P.~Devanbu, A.~Oliveira, V.~Filkov, and B.~Ray.
\newblock Assert use in github projects.
\newblock In {\em Proceedings of the 37th International Conference on Software
  Engineering - Volume 1}, ICSE '15, pages 755--766, Piscataway, NJ, USA, 2015.
  IEEE Press.

\bibitem{icse17type}
M.~Coblenz, W.~Nelson, J.~Aldrich, B.~Myers, and J.~Sunshine.
\newblock Glacier: Transitive class immutability for java.
\newblock In {\em Proceedings of the 39th International Conference on Software
  Engineering}, ICSE '17, pages 496--506, Piscataway, NJ, USA, 2017. IEEE
  Press.

\bibitem{modelimpact}
J.~Cohen and P.~Cohen.
\newblock {\em Applied multiple regression/correlation analysis for the
  behavioral sciences}.
\newblock Lawrence Erlbaum Associates, 1975.

\bibitem{premregressionmodel}
P.~Devanbu, T.~Zimmermann, and C.~Bird.
\newblock Belief \&\#38; evidence in empirical software engineering.
\newblock In {\em Proceedings of the 38th International Conference on Software
  Engineering}, ICSE '16, pages 108--119, New York, NY, USA, 2016. ACM.

\bibitem{type-checker}
W.~Dietl, S.~Dietzel, M.~D. Ernst, K.~Mu\c{s}lu, and T.~W. Schiller.
\newblock Building and using pluggable type-checkers.
\newblock In {\em Proceedings of the 33rd International Conference on Software
  Engineering}, ICSE '11, pages 681--690, New York, NY, USA, 2011. ACM.

\bibitem{ecoopcontract}
J.~Dietrich, D.~J. Pearce, K.~Jezek, and P.~Brada.
\newblock Contracts in the wild: A study of java programs.
\newblock In {\em LIPIcs-Leibniz International Proceedings in Informatics},
  volume~74. Schloss Dagstuhl-Leibniz-Zentrum fuer Informatik, 2017.

\bibitem{annotationicse}
R.~Dyer, H.~Rajan, H.~A. Nguyen, and T.~N. Nguyen.
\newblock Mining billions of ast nodes to study actual and potential usage of
  java language features.
\newblock In {\em Proceedings of the 36th International Conference on Software
  Engineering}, ICSE 2014, pages 779--790, New York, NY, USA, 2014. ACM.

\bibitem{classsize}
K.~E. Emam, S.~Benlarbi, N.~Goel, and S.~N. Rai.
\newblock The confounding effect of class size on the validity of
  object-oriented metrics.
\newblock {\em IEEE Transactions on Software Engineering}, 27(7):630--650, Jul
  2001.

\bibitem{Ernst2016}
M.~D. Ernst, A.~Lovato, D.~Macedonio, F.~Spoto, and J.~Thaine.
\newblock Locking discipline inference and checking.
\newblock In {\em ICSE 2016, Proceedings of the 38th International Conference
  on Software Engineering}, pages 1133--1144, Austin, TX, USA, May 2016.

\bibitem{cannotation}
D.~Evans.
\newblock Static detection of dynamic memory errors.
\newblock In {\em Proceedings of the ACM SIGPLAN 1996 Conference on Programming
  Language Design and Implementation}, PLDI '96, pages 44--53, New York, NY,
  USA, 1996. ACM.

\bibitem{gumtree}
J.~Falleri, F.~Morandat, X.~Blanc, M.~Martinez, and M.~Monperrus.
\newblock Fine-grained and accurate source code differencing.
\newblock In {\em {ACM/IEEE} International Conference on Automated Software
  Engineering, {ASE} '14, Vasteras, Sweden - September 15 - 19, 2014}, pages
  313--324, 2014.

\bibitem{ESCChecker}
C.~Flanagan, K.~R.~M. Leino, M.~Lillibridge, G.~Nelson, J.~B. Saxe, and
  R.~Stata.
\newblock Extended static checking for java.
\newblock In {\em Proceedings of the ACM SIGPLAN 2002 Conference on Programming
  Language Design and Implementation}, PLDI '02, pages 234--245, New York, NY,
  USA, 2002. ACM.

\bibitem{JLS2}
J.~Gosling, B.~Joy, G.~Steele, and G.~Bracha.
\newblock {\em Java Language Specification, Second Edition: The Java Series}.
\newblock Addison-Wesley Longman Publishing Co., Inc., Boston, MA, USA, 2nd
  edition, 2000.

\bibitem{JLS3}
J.~Gosling, B.~Joy, G.~Steele, and G.~Bracha.
\newblock {\em Java(TM) Language Specification, The (3rd Edition) (Java
  (Addison-Wesley))}.
\newblock Addison-Wesley Professional, 2005.

\bibitem{Java8}
J.~Gosling, B.~Joy, G.~L. Steele, G.~Bracha, and A.~Buckley.
\newblock {\em The Java Language Specification, Java SE 8 Edition}.
\newblock Addison-Wesley Professional, 1st edition, 2014.

\bibitem{ESEMJava}
M.~Grechanik, C.~McMillan, L.~DeFerrari, M.~Comi, S.~Crespi, D.~Poshyvanyk,
  C.~Fu, Q.~Xie, and C.~Ghezzi.
\newblock An empirical investigation into a large-scale java open source code
  repository.
\newblock In {\em Proceedings of the 2010 ACM-IEEE International Symposium on
  Empirical Software Engineering and Measurement}, ESEM '10, pages 11:1--11:10,
  New York, NY, USA, 2010. ACM.

\bibitem{annotationoveruse}
E.~Guerra.
\newblock Design patterns for annotation-based apis.
\newblock In {\em Proceedings of the 11th Latin-American Conference on Pattern
  Languages of Programming}, SugarLoafPLoP '16, pages 9:1--9:14, USA, 2016. The
  Hillside Group.

\bibitem{ICSE16energy}
S.~Hasan, Z.~King, M.~Hafiz, M.~Sayagh, B.~Adams, and A.~Hindle.
\newblock Energy profiles of java collections classes.
\newblock In {\em Proceedings of the 38th International Conference on Software
  Engineering}, ICSE '16, pages 225--236, New York, NY, USA, 2016. ACM.

\bibitem{OOSPLAGenerics}
M.~Hoppe and S.~Hanenberg.
\newblock Do developers benefit from generic types?: An empirical comparison of
  generic and raw types in java.
\newblock In {\em Proceedings of the 2013 ACM SIGPLAN International Conference
  on Object Oriented Programming Systems Languages \&\#38; Applications},
  OOPSLA '13, pages 457--474, New York, NY, USA, 2013. ACM.

\bibitem{hurdlemodel}
M.-C. Hu, M.~Pavlicova, and E.~V. Nunes.
\newblock Zero-inflated and hurdle models of count data with extra zeros:
  examples from an hiv-risk reduction intervention trial.
\newblock {\em The American journal of drug and alcohol abuse}, 37(5):367--375,
  2011.

\bibitem{oopslatype}
W.~Huang, A.~Milanova, W.~Dietl, and M.~D. Ernst.
\newblock Reim \&\#38; reiminfer: Checking and inference of reference
  immutability and method purity.
\newblock In {\em Proceedings of the ACM International Conference on Object
  Oriented Programming Systems Languages and Applications}, OOPSLA '12, pages
  879--896, New York, NY, USA, 2012. ACM.

\bibitem{RSS}
S.~G.~P. John O.~Rawlings and D.~A. Dickey.
\newblock {\em Applied RegressionAnalysis:A Research Tool}.
\newblock Springer-Verlag New York.

\bibitem{tutoriallombok}
J.~Josh.
\newblock Project lombok: Clean, concise java code.
\newblock \url{https://www.oracle.com/corporate/features/project-lombok.html},
  2017.

\bibitem{annotationoveruseweb1}
T.~Jungle.
\newblock Annotations and its benefits in java.
\newblock
  \url{http://technicaljungle.com/annotations-in-java-intro-benefits-avoidance/},
  2018.

\bibitem{Kalliamvakou:2016:ISP:2992358.2992445}
E.~Kalliamvakou, G.~Gousios, K.~Blincoe, L.~Singer, D.~M. German, and
  D.~Damian.
\newblock An in-depth study of the promises and perils of mining github.
\newblock {\em Empirical Softw. Engg.}, 21(5):2035--2071, Oct. 2016.

\bibitem{Alloy}
S.~Khurshid, D.~Marinov, and D.~Jackson.
\newblock An analyzable annotation language.
\newblock In {\em Proceedings of the 17th ACM SIGPLAN Conference on
  Object-oriented Programming, Systems, Languages, and Applications}, OOPSLA
  '02, pages 231--245, New York, NY, USA, 2002. ACM.

\bibitem{commitclassify4}
P.~S. Kochhar and D.~Lo.
\newblock Revisiting assert use in github projects.
\newblock In {\em Proceedings of the 21st International Conference on
  Evaluation and Assessment in Software Engineering}, EASE'17, pages 298--307,
  New York, NY, USA, 2017. ACM.

\bibitem{icsereflection}
D.~Landman, A.~Serebrenik, and J.~J. Vinju.
\newblock Challenges for static analysis of java reflection - literature review
  and empirical study.
\newblock In {\em 2017 IEEE/ACM 39th International Conference on Software
  Engineering (ICSE)}, pages 507--518, May 2017.

\bibitem{anna}
D.~Luckham and F.~W. Henke.
\newblock An overview of anna - a specification language for ada.
\newblock Technical report, Stanford, CA, USA, 1984.

\bibitem{nlptool}
C.~D. Manning, M.~Surdeanu, J.~Bauer, J.~Finkel, S.~J. Bethard, and
  D.~McClosky.
\newblock The stanford corenlp natural language processing toolkit.
\newblock In {\em Proceedings of 52nd Annual Meeting of the Association for
  Computational Linguistics: System Demonstrations}, 2014.

\bibitem{oopsla2017}
D.~Mazinanian, A.~Ketkar, N.~Tsantalis, and D.~Dig.
\newblock Understanding the use of lambda expressions in java.
\newblock {\em Proc. ACM Program. Lang.}, 1(OOPSLA):85:1--85:31, Oct. 2017.

\bibitem{tutorialmsdn}
Microsoft.
\newblock Sal annotations.
\newblock \url{https://msdn.microsoft.com/en-us/library/ms235402.aspx}, 2015.

\bibitem{icsmbug}
A.~Mockus and L.~G. Votta.
\newblock Identifying reasons for software changes using historic databases.
\newblock In {\em Proceedings of the International Conference on Software
  Maintenance (ICSM'00)}, ICSM '00, pages 120--, Washington, DC, USA, 2000.
  IEEE Computer Society.

\bibitem{icse05codechurn}
N.~Nagappan and T.~Ball.
\newblock Use of relative code churn measures to predict system defect density.
\newblock In {\em Proceedings of the 27th International Conference on Software
  Engineering}, ICSE '05, pages 284--292, New York, NY, USA, 2005. ACM.

\bibitem{annotationplace}
Oracle.
\newblock Annotation type suppresswarnings.
\newblock
  \url{https://docs.oracle.com/javase/7/docs/api/java/lang/SuppressWarnings.html},
  2017.

\bibitem{tutorial}
Oracle.
\newblock Lesson: Annotations.
\newblock \url{https://docs.oracle.com/javase/tutorial/java/annotations/},
  2017.

\bibitem{tutorialoracleannotation}
Oracle.
\newblock Annotations.
\newblock
  \url{https://docs.oracle.com/javase/7/docs/technotes/guides/language/annotations.html},
  2018.

\bibitem{annotationoveruseweb3}
S.~Overflow.
\newblock Arguments against annotations.
\newblock
  \url{https://stackoverflow.com/questions/1675610/arguments-against-annotations},
  2018.

\bibitem{isstatypechecker}
M.~M. Papi, M.~Ali, T.~L. Correa, Jr., J.~H. Perkins, and M.~D. Ernst.
\newblock Practical pluggable types for java.
\newblock In {\em Proceedings of the 2008 International Symposium on Software
  Testing and Analysis}, ISSTA '08, pages 201--212, New York, NY, USA, 2008.
  ACM.

\bibitem{genericsemse}
C.~Parnin, C.~Bird, and E.~Murphy-Hill.
\newblock Adoption and use of java generics.
\newblock {\em Empirical Softw. Engg.}, 18(6):1047--1089, Dec. 2013.

\bibitem{spoonlib}
R.~Pawlak, M.~Monperrus, N.~Petitprez, C.~Noguera, and L.~Seinturier.
\newblock Spoon: A library for implementing analyses and transformations of
  java source code.
\newblock {\em Software: Practice and Experience}, 46:1155--1179, 2015.

\bibitem{ownership1}
F.~Rahman and P.~Devanbu.
\newblock Ownership, experience and defects: A fine-grained study of
  authorship.
\newblock In {\em Proceedings of the 33rd International Conference on Software
  Engineering}, ICSE '11, pages 491--500, New York, NY, USA, 2011. ACM.

\bibitem{commitclassify2}
B.~Ray, V.~Hellendoorn, Z.~Tu, C.~Nguyen, S.~Godhane, A.~Bacchelli, and
  P.~Devanbu.
\newblock On the" naturalness" of buggy code.
\newblock ICSE '16. ACM, 2016.

\bibitem{commitclassify3}
B.~Ray, D.~Posnett, V.~Filkov, and P.~Devanbu.
\newblock A large scale study of programming languages and code quality in
  github.
\newblock In {\em Proceedings of the ACM SIGSOFT 22nd International Symposium
  on the Foundations of Software Engineering}, FSE '14. ACM, 2014.

\bibitem{annotationoveruseweb2}
G.~Riegler.
\newblock Be a better developer.
\newblock
  \url{http://www.beabetterdeveloper.com/2013/12/an-annotation-nightmare.html},
  2018.

\bibitem{annotationseke}
H.~Rocha and M.~T. Valente.
\newblock How annotations are used in java: An empirical study.
\newblock In {\em SEKE}, pages 426--431. Knowledge Systems Institute Graduate
  School, 2011.

\bibitem{Tanannotation}
L.~Tan, Y.~Zhou, and Y.~Padioleau.
\newblock acomment: Mining annotations from comments and code to detect
  interrupt related concurrency bugs.
\newblock In {\em Proceedings of the 33rd International Conference on Software
  Engineering}, ICSE '11, pages 11--20, New York, NY, USA, 2011. ACM.

\bibitem{JavaField}
E.~Tempero.
\newblock How fields are used in java: An empirical study.
\newblock In {\em 2009 Australian Software Engineering Conference}, pages
  91--100, April 2009.

\bibitem{Javaheritance}
E.~Tempero, J.~Noble, and H.~Melton.
\newblock How do java programs use inheritance? an empirical study of
  inheritance in java software.
\newblock In {\em Proceedings of the 22Nd European Conference on
  Object-Oriented Programming}, ECOOP '08, pages 667--691, Berlin, Heidelberg,
  2008. Springer-Verlag.

\bibitem{OOPSLA15}
H.~Zhang, Z.~Chu, B.~C. d.~S. Oliveira, and T.~v.~d. Storm.
\newblock Scrap your boilerplate with object algebras.
\newblock In {\em Proceedings of the 2015 ACM SIGPLAN International Conference
  on Object-Oriented Programming, Systems, Languages, and Applications}, OOPSLA
  2015, pages 127--146, New York, NY, USA, 2015. ACM.

\end{thebibliography}

\end{document}